\documentclass[reprint,amsfonts, amssymb, amsmath,  showkeys,prx, superscriptaddress, twocolumn,longbibliography,nofootinbib]{revtex4-2}
\usepackage{float}
\makeatletter
\let\newfloat\newfloat@ltx
\makeatother
\usepackage[english]{babel}
\usepackage{algorithmic}
\usepackage[utf8]{inputenc}
\usepackage{graphics}
\usepackage{selinput}
\usepackage[normalem]{ulem}
\usepackage[shortlabels]{enumitem}

\usepackage{braket}
\usepackage{amsthm}
\usepackage{mathtools}
\usepackage{physics}
\usepackage{xcolor}
\usepackage{graphicx}
\usepackage[left=16mm,right=16mm,top=35mm,columnsep=15pt]{geometry} 
\usepackage{adjustbox}
\usepackage{placeins}
\usepackage[T1]{fontenc}
\usepackage{lipsum}
\usepackage{csquotes}
\usepackage{bm}

\usepackage[linesnumbered,ruled,vlined]{algorithm2e}
\SetKwInput{kwInit}{Init}

\def\HC{\mathcal{H}}

\def\ad{^{\dagger}}

\newcommand{\fsnull}[1]{}
\newcommand{\old}[1]{}

\usepackage[makeroom]{cancel}
\usepackage[toc,page]{appendix}
\usepackage[colorlinks=true,citecolor=blue,linkcolor=magenta]{hyperref}

\usepackage{tikz}
\tikzset{every picture/.style=remember picture}

\usepackage[utf8]{inputenc}
\usepackage{graphicx}
\usepackage{xcolor}
\usepackage{amsmath}
\usepackage{amsthm}
\usepackage{bm}
\usepackage{bbm}
\usepackage{comment}
\usepackage{appendix}
\usepackage{mathdots}
\usepackage{lipsum}
\usepackage{verbatim}
\usepackage{natbib}
\usepackage{nccmath}
\usepackage{amsfonts} 

\newcommand{\dya}[1]{\ket{#1}\!\bra{#1}}

\newcommand{\poly}{\operatorname{poly}}

\newcommand{\BC}{\mathcal{B}}

\newcommand{\GC}{\mathcal{G}}

\newcommand{\NC}{\mathcal{N}}
\newcommand{\OC}{\mathcal{O}}

\newcommand{\TC}{\mathcal{T}}

\renewcommand{\geq}{\geqslant}
\renewcommand{\leq}{\leqslant}

\renewcommand{\vec}[1]{\boldsymbol{#1}}  

\newcommand*{\id}{\openone}

\newcommand{\bs}{\textsf{BS}}

\def\be{\begin{equation}}
\def\ee{\end{equation}}
\def\bs{\begin{split}}
\def\e{\end{split}}
\def\ba{\begin{eqnarray}}
\def\bea{\begin{eqnarray}}

\def\tea{\end{eqnarray}}
\def\ea{\end{eqnarray}}
\def\eea{\end{eqnarray}}

\newtheorem{theorem}{Theorem}
\newtheorem{lemma}{Lemma}

\newtheorem{proposition}{Proposition}

\newcommand{\ingr}[2]{\begin{matrix}\includegraphics[height=#1 cm]{figures/inline/#2}\end{matrix}}
\newcommand{\ingrp}[2]{\begin{matrix}\includegraphics[height=#1 cm]{figures/inline2/#2}\end{matrix}}

\usepackage{amssymb}
\usepackage{dsfont}

\def\be{\begin{equation}}
\def\te{\end{equation}}
\def\ee{\end{equation}}
\def\ba{\begin{eqnarray}}
\def\bea{\begin{eqnarray}}

\def\tea{\end{eqnarray}}
\def\ea{\end{eqnarray}}
\def\eea{\end{eqnarray}}

\begin{document}

\title{Computing exact moments of local random quantum circuits via tensor networks}

\author{Paolo Braccia}
\affiliation{Theoretical Division, Los Alamos National Laboratory, Los Alamos, NM 87545, USA}

\author{Pablo Bermejo}
\affiliation{Information Sciences, Los Alamos National Laboratory, Los Alamos, NM 87545, USA}
\affiliation{Donostia International Physics Center, Paseo Manuel de Lardizabal 4, E-20018 San Sebasti\'an, Spain}
\author{Lukasz Cincio}
\affiliation{Theoretical Division, Los Alamos National Laboratory, Los Alamos, NM 87545, USA}

\author{M. Cerezo}
\thanks{cerezo@lanl.gov}
\affiliation{Information Sciences, Los Alamos National Laboratory, Los Alamos, NM 87545, USA}

\begin{abstract}

A basic primitive in quantum  information is the computation of the moments $\mathbb{E}_U[\Tr[U\rho U\ad O]^t]$. These describe the distribution of expectation values  obtained by  sending a state $\rho$ through a random unitary $U$, sampled from some distribution, and measuring the observable $O$. While the exact calculation of these moments is generally hard, if $U$ is composed of local random gates, one can estimate $\mathbb{E}_U[\Tr[U\rho U\ad O]^t]$ by performing Monte Carlo simulations of a Markov chain-like process. However, this approach can require a prohibitively large number of samples, or suffer from the sign problem. In this work, we instead propose to estimate the moments via tensor networks, where the local gates moment operators are mapped to small dimensional tensors acting on their local commutant bases. By leveraging representation theoretical tools, we study the local tensor dimension and we provide bounds for the bond dimension of the  matrix product states arising from deep circuits. We compare our techniques against Monte Carlo simulations, showing that we can significantly out-perform them. Then, we showcase how tensor networks can exactly compute  the second moment  when $U$ is a quantum neural network acting on thousands of qubits and having thousands of gates.   To finish, we numerically study the anticoncentration phenomena of circuits with orthogonal random gates, a task which cannot be studied via Monte Carlo due to sign problems.

\end{abstract}

    \maketitle

\section{Introduction}

Computing the moments of expectation values measured at the output of random quantum circuits has become an important task in quantum information sciences. For instance, their analysis can help us determine conditions leading to non-classical simulability and exponential quantum advantage~\cite{boixo2018characterizing,arute2019quantum,wu2021strong,dalzell2022randomquantum,oszmaniec2022fermion,huang2021provably}, the onset of quantum chaos~\cite{nahum2017quantum,von2018operator,nahum2018operator}, and the presence of local minima and barren plateaus in variational quantum computing~ \cite{mcclean2018barren,cerezo2020cost,pesah2020absence,napp2022quantifying,ragone2023unified,fontana2023theadjoint,diaz2023showcasing,letcher2023tight,thanaslip2021subtleties,anschuetz2022quantum,anschuetz2021critical,monbroussou2023trainability}. When the circuit is sampled from a compact unitary group, one can leverage tools from Weingarten calculus~\cite{mele2023introduction,ragone2022representation} to exactly evaluate these moments, even being able to compute them asymptotically~\cite{garcia2023deep}. However, if the unitaries $U$ are sampled from a generic distribution, the calculation quickly becomes extremely hard and intractable. 

To make the problem more manageable, researchers have studied circuits which have additional structure to them. One of the most physically motivated  assumptions is that $U$ is composed of local Haar random gates~\cite{hayden2007black,sekino2008fast,brown2012scrambling,lashkari2013towards,hosur2016chaos,nahum2018operator,von2018operator}. In this framework, one can map  the problem of computing the moment to a Markov chain-like process, which enables the use of tools from classical statistical mechanics~\cite{hunter2019unitary,barak2020spoofing,napp2022quantifying,harrow2018approximate,letcher2023tight,hayden2016holographic,nahum2018operator,hunter2019unitary,harrow2018approximate}. This approach has shown to be incredibly successful, and it particularly excels at producing upper and lower bounds for the moments. However, there is still room for improvement. In particular, for most of the aforementioned techniques to work, it is not always sufficient to consider that the unitary is composed of local random gates, as other assumptions are usually needed. For instance, one could require that the gates are sampled from the same group, or that the circuit gates have either a very regular arrangement or a very random one. Moreover, one could also be interested in computing the exact value of the moments, and not their bound. While some of these limitations can be overcome by using Monte Carlo (MC) sampling of the Markov chain-like process~\cite{napp2022quantifying}, this approach can still require a prohibitively large number of samples due to its additive error, or suffer from the sign problem~\cite{pan2022sign,mi2021information}.

\begin{figure*}[t]
    \centering
    \includegraphics[width=1\linewidth]{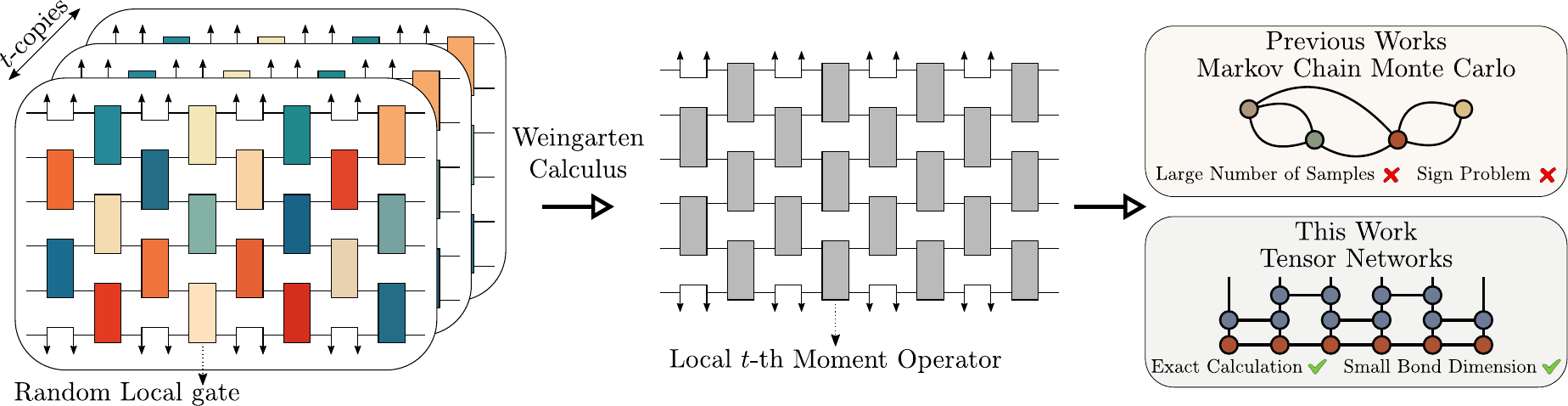}
    \caption{Schematic representation of our main results. Our ultimate goal is to estimate expectation values of the form $\mathbb{E}_U[\Tr[U\rho U\ad O]^t]$, where $\rho$ is a quantum state, $U$ a quantum circuit composed of local random gates, and $O$ some observable. By leveraging tools from the Weingarten calculus we can map this problem  to that of evaluating the inner product $
   \mathbb{E}_U[\Tr[U \rho U\ad O]^t]= \langle\langle \rho^{\otimes t}|\widehat{\tau}^{(t)}|O^{\otimes t}\rangle\rangle$, where $|\rho^{\otimes t}\rangle\rangle$ and $|O^{\otimes t}\rangle\rangle$ are respectively the vectorized $t$-th fold tensor product of the initial state and measurement, and where $\widehat{\tau}^{(t)}$ is the product of the local gates $t$-th moment operators. Previous works have proposed to compute this inner product by interpreting $\widehat{\tau}^{(t)}$ as a Markov chain-like process and using  MC sampling techniques. In this work, we instead evaluate this quantity by expressing $|\rho^{\otimes t}\rangle\rangle$ and $|O^{\otimes t}\rangle\rangle$ as MPSs, and  $\widehat{\tau}^{(t)}$ as a TN with local gates.    }
    \label{fig:schematic}
\end{figure*}

In this work we propose using Tensor Networks (TN) to compute the moments of the random quantum circuit's expectation values. At its core, our work leverages the vectorization picture to represent the circuit's initial state and the measurement operator $t$-th fold products as Matrix Product State (MPS)~\cite{orus2014practical,biamonte2017tensor} and the local unitaries moment matrices as gates arising from the projectors between their local commutants (see Fig.~\ref{fig:schematic}). This formalism is quite flexible, being applicable to general gate topologies, as well as circuits where the local gates are sampled from different groups. Of course, we are not claiming that the tensor contractions will always be efficient, since their difficulty will depend on several factors such as the gate's topology, and the local dimensions of the tensors. Still, our approach has several advantages over MC techniques. For example TNs do not suffer from sign problems, nor require sampling.  In addition, we also make theoretical contributions by leveraging tools from representation theory to compute the local dimension of the  TNs, as well as provide bounds for the maximum bond dimension of MPSs arising from deep circuits. Moreover, we showcase our methods by comparing them against  MC sampling, revealing that we can indeed compute the moments with much higher precision when using TN. We then present large scale numerical experiments  which illustrate the power of TNs for studying the variance of expectation value. Finally, we show that having access to the MPSs representation of the initial state (or measurement) as it evolves through the TN moment gates enables a new dimension of analysis, such as the study of this vector's entropic properties.

\section{Framework}

In what follows we will consider a random unitary quantum circuit $U$ acting on an $n$-qudit Hilbert space $\HC=(\mathbb{C}^d)^{\otimes n}$, where $d$ is the dimensionality of a single qudit. We further assume that the circuit takes the form $U=\prod_{l=1}^L  U_{\gamma_l}$, i.e., that it is composed of $L$ local gates acting on  $k_l\leq k$ qudits according to some  topology $\TC=\{\gamma_l\}_{l=1}^L$. Here, each  $\gamma_l\in[n]^{\times k_l}$ is a tuple of (non-repeating) $k_l$ indexes ranging between $1$ and $n$ which determine the set of qudits that $U_{\gamma_l}$ acts on. For instance, if the circuit is a one-dimensional ansatz acting on alternating pairs of qudits (see Fig.~\ref{fig:schematic}), then we would have $\TC=\{(1,2),(3,4),\ldots (n-1,n),(2,3),\ldots\}$.  Moreover, we will assume that each $U_{\gamma_l}$ belongs to some unitary Lie group $G_l\subseteq U(d^k)$. 

Next, we will focus on studying moments of expectation values of the form 
\begin{equation}
    \ell_U(\rho,O)=\Tr[U\rho U\ad O ]\,,
\end{equation}
which correspond to sending an input $n$-qudit state $\rho$ though $U$ and measuring the expectation value of a Hermitian observable $O$ at the circuit's output. Here, by moments we mean analyzing quantities such as
\begin{equation}\label{eq:moments}
    \mathbb{E}_U[\ell_U(\rho,O)]\,, \quad \mathbb{E}_U[(\ell_U(\rho,O))^2]\,, 
\end{equation}
or more generally $\mathbb{E}_U[(\ell_U(\rho,O))^t]$\footnote{We note that while not explicitly stated, our methods can be trivially extended to compute more general moments such as $\mathbb{E}_U[\prod_{\eta=1}^t\Tr[U\rho_\eta U\ad O_\eta ]]$. } for some $t\in\mathbb{N}$. In the previous, the expectation value is taken over the set of unitaries obtained by sampling  each  local $U_{\gamma_l}$  independently and identically distributed (i.i.d.)  according to the Haar measure $d\mu_l$ over $G_l$. As such, we can express $\mathbb{E}_U=\mathbb{E}_{G_L}\cdots \mathbb{E}_{G_1} $, with $\mathbb{E}_{G}[f(U)]=\int_{G}d\mu(U)f(U)$ the Haar integral over the group $G$.

\section{Connection to the literature}\label{sec:connection_to_literature}

Before proceeding to the description of our results, we find it important to compare our methods with other techniques that exist in the literature. In particular, given the pervasiveness of quantities such as those in Eq.~\eqref{eq:moments}, their computation has received considerable attention. The simplest  approach to computing these expectation values is by running the circuit $N$ times (sampling random local unitaries each time) and computing the sample mean of the set of observations. While such approach is by far the simplest, it has the significant --and often overlooked-- disadvantage that the ensuing expectation values come with a statistical uncertainty of $1/\sqrt{N}$. As such, computing the moments to a large precision requires a large number of circuit repetitions.

A more principled theoretical approach to computing the values in Eq.~\eqref{eq:moments} has been taken in~\cite{hunter2019unitary,cerezo2020cost,pesah2020absence,barak2020spoofing,dalzell2022randomquantum,napp2022quantifying,harrow2018approximate,letcher2023tight} where the authors leverage tools from Weingarten calculus to analytically compute averages over each local Haar random gate. While one might try to brute-force this problem (e.g., via symbolic integration~\cite{fukuda2019rtni}), the number of terms one needs to keep track of grows exponentially with $t$ and with the number of gates. However, as  further detailed below, Refs.~\cite{hunter2019unitary,cerezo2020cost,pesah2020absence,barak2020spoofing,dalzell2022randomquantum,napp2022quantifying,harrow2018approximate,letcher2023tight} mitigate this issue by mapping the problem of evaluating $\mathbb{E}_U[(\ell_U(\rho,O))^t]$ to a Markov chain-like problem, therefore enabling the use of classical statistical mechanics~\cite{hayden2016holographic,nahum2018operator,hunter2019unitary,harrow2018approximate,dalzell2022randomquantum,napp2022quantifying}. In particular,  these works  consider exclusively circuits where all the local unitaries are sampled i.i.d. from the same fundamental representation of a unitary group, and where the circuit's topology is somewhat structured. Moreover, their goal  is to ultimately provide \textit{upper and lower bounds} for the expectation value's moments. While such bounds can indeed be valuable, they can be quite loose. 

It is also important to note that for most cases considered in the literature, e.g., the expectation values  with $t=1$ and $t=2$ as in Eq.~\eqref{eq:moments}, one can use MC methods to approximate the expectation values through the ensuing Markov chain-like approach~\cite{napp2022quantifying}. While MC techniques can lead to good approximations of the moments, they have two critical issues. First, one needs to run the MC sampling algorithm a finite number of times, leading to an additive error. Since the  expectation values $\mathbb{E}_U[(\ell_U(\rho,O))^t]$ can be exponentially small quantities, one could require a  prohibitively large number of samples for their approximation. Second, and most importantly, the weights of the configuration can be negative, meaning that one can encounter the sign problem, which  increases exponentially the computational complexity of the MC simulation~\cite{pan2022sign,mi2021information}.

In this work, we will follow a similar approach to the one in~\cite{hunter2019unitary,cerezo2020cost,pesah2020absence,barak2020spoofing,dalzell2022randomquantum,napp2022quantifying,harrow2018approximate,letcher2023tight}, as we make use of Weingarten to map the problem of calculating $\mathbb{E}_U[(\ell_U(\rho,O))^t]$ to that of evolving a state through a Markov chain-like process matrix. However,  unlike previous  approaches which leverage statistical mechanics or MC simulations, we will here   present a TN method which allows us to exactly obtain the expectation values. Our techniques instead can be  applied under quite general conditions, including the case when the local unitaries are sampled from subgroups of the unitary (such as the orthogonal group) and combinations thereof, as well as when the circuit topology is arbitrary.  Moreover, as we explicitly show in our numerical experiments section, the TN techniques introduced here can be employed in situations where a sign problem would arise for MC techniques.

\section{From Weingarten calculus to tensor networks}

\subsection{Basics of Weingarten calculus}

To begin, we recall a few basic concepts from Weingarten calculus. We refer the interested reader to~\cite{mele2023introduction,ragone2022representation,garcia2023deep} for a modern and detailed description of the tools used here. 

Given a compact unitary Lie group $G$ with Haar measure $d\mu$ acting on a finite-dimensional Hilbert space $\HC$, we define the $t$-th fold twirl map $\tau_{G}^{(t)}:\BC(\HC^{\otimes t })\rightarrow \BC(\HC^{\otimes t })$ as 
\begin{equation}\label{eq:t_twirl}
    \tau_{G}^{(t)}(X) = \int_G d\mu(U) U^{\otimes t}X (U^\dagger)^{\otimes t}\,,
\end{equation}
for any $X\in \BC(\HC^{\otimes t })$. Here $\BC$ denotes the set of bounded operators acting on a given vector space. The importance of the twirl spans from the fact that any expectation value of the form $\mathbb{E}_G[\Tr[U \rho U\ad O]^t]$ can be written as
\small
\begin{equation}
   \mathbb{E}_G\left[\Tr[U \rho U\ad O]^t\right]= \Tr[\tau_{G}^{(t)}(\rho^{\otimes t}) O^{\otimes t}]=\Tr[ \rho^{\otimes t} \tau_{G}^{(t)}(O^{\otimes t})]\,,\nonumber
\end{equation}
\normalsize
where we have used the linearity of the trace and the invariance of the Haar measure. 

Then, we find it convenient to recall the vectorization 
 formalism, in which an operator in $\BC(\HC^{\otimes t})$ is mapped to a vector in $\HC^{\otimes t}\otimes (\HC^*)^{\otimes t}$ while a channel from $\BC(\HC^{\otimes t })$ to $\BC(\HC^{\otimes t })$ is mapped to a matrix in $\BC(\HC^{\otimes t}\otimes (\HC^*)^{\otimes t})$. Specifically, given some operator $X=\sum_{i,j=1}^{d^t} c_{ij}|i\rangle\langle j|$, its vectorized form is $|X\rangle \rangle=\sum_{i,j=1}^{d^t} c_{ij}|i\rangle\otimes | j\rangle$, while given a channel $\Phi(X)=\sum_{\nu=1}^{d^{2t}} K_\nu X J_\nu\ad$, we obtain $\widehat{\Phi}=\sum_{\nu=1}^{d^{2t}}K_\nu \otimes  J_\nu^*$. In this picture, we can verify that since $\langle\langle Y|\widehat{\Phi}|X\rangle \rangle=\Tr[Y\ad \Phi(X)]$, then 
 \begin{equation}\label{eq:expectation-vectorized}
   \mathbb{E}_G\left[\Tr[U \rho U\ad O]^t\right]= \langle\langle \rho^{\otimes t}|\widehat{\tau}_{G}^{(t)}|O^{\otimes t}\rangle \rangle \,,
\end{equation}
where the Hermitian operator $\widehat{\tau}_{G}^{(t)}\in\BC(\HC^{\otimes t}\otimes (\HC^*)^{\otimes t})$, defined as
\begin{equation}
    \widehat{\tau}_{G}^{(t)}= \int_G d\mu(U) U^{\otimes t}\otimes (U^*)^{\otimes t}
    \widehat{\tau}_{G}^{(t)}= \int_G d\mu(U) U^{\otimes t}\otimes (U^*)^{\otimes t}
\end{equation}
is known as the $t$-th moment operator.

It is well known that the twirl map is a projection onto  the $t$-th order commutant of $G$, i.e., the operator vector  space $\textrm{comm}^{(t)}(G)=\{M\in\BC(\HC^{\otimes t })\,|\, [M,U^{\otimes t}]=0\}$, of dimension $d_{G,t}=\dim(\textrm{comm}^{(t)}(G))$. As such, given a basis $\{P_\mu\}_{\mu=1}^{d_{G,t}}$ of $\textrm{comm}^{(t)}(G)$, one can express the twirl operator as 
\begin{equation}\label{eq:twirl-commutant}
    \tau_{G}^{(t)}(X) = \sum_{\mu,\nu=1}^{d_{G,t}} (W_{G,t}^{-1})_{\nu\mu} \Tr[P_\mu X] P_\nu.
\end{equation}
Here $W^{-1}_{G,t}$ is known as the Weingarten Matrix, and is the inverse\footnote{Or pseudo-inverse if $W$ is singular.} of the Gram matrix $W$ of the commutant's basis (under the Hilbert-Schmidt inner product). That is, $W_{G,t}$ is a matrix with entries $(W_{G,t})_{\nu\mu}=\Tr[P_\nu \ad P_\mu]$. In the vectorization picture, we have that 
\begin{equation}\label{eq:twirl-vectorized}
    \widehat{\tau}_{G}^{(t)}=\sum_{\mu,\nu=1}^{d_{G,t}} (W_{G,t}^{-1})_{\nu\mu} | P_\nu\rangle\rangle\langle\langle P_\mu|\,.
\end{equation}

\subsection{Mapping to a tensor network problem}

Using the previous tools and techniques, we can thus express 
\begin{align}\label{eq:momenti-inner-prod}
\mathbb{E}_U[(\ell_U(\rho,O))^t]&=\langle\langle \rho^{\otimes t}|\prod_{l=1}^L\widehat{\tau}_{G_l}^{(t)}|O^{\otimes t}\rangle \rangle\\
&=\langle\langle \rho^{\otimes t}|\widehat{\tau}^{(t)}|O^{\otimes t}\rangle \rangle\,.
\end{align}
where we defined $\widehat{\tau}^{(t)}=\prod_{l=1}^L\widehat{\tau}_{G_l}^{(t)}$. Equation~\eqref{eq:momenti-inner-prod} shows that in order to compute the expectation value, we need to compute the inner product between two vectors ($|\rho^{\otimes t}\rangle \rangle$ and $|O^{\otimes t}\rangle \rangle$) and the $d^{2tn}\times d^{2tn}$ matrix $\widehat{\tau}^{(t)}$. Clearly, the evaluation of this expectation value becomes rapidly intractable as its dimension increases exponentially with the number of qudits $n$, and the moment's order $t$. However, since the gates are local, it is completely unnecessary to work with such large matrices, and the problem's dimension can be reduced to a much smaller problem size~\cite{dalzell2022randomquantum,napp2022quantifying}. 

To understand the previous let us note that   $\widehat{\tau}^{(t)}$ is a product of the local moment superoperators for each gate. Then, we can see  by 
explicitly expanding the product of two adjacent moment operators that 
\begin{align}
\widehat{\tau}_{G_{l+1}}^{(t)}\widehat{\tau}_{G_l}^{(t)}=\sum_{\mu_{l+1},\nu_{l+1}}\sum_{\mu_l,\nu_l}&(W_{G_l,t}^{-1})_{\nu_l\mu_l}(W_{G_{l+1},t}^{-1})_{\nu_{l+1}\mu_{l+1}}\nonumber\\
&\times | P_{\nu_{l+1}}\rangle\rangle\langle\langle P_{\mu_{l+1}}
 | P_{\nu_l}\rangle\rangle\langle\langle P_{\mu_l}|\nonumber\,.
\end{align}
That is, the product of two moment operators can be understood as a map between the basis elements of the $t$-th order commutants of the relative gates. As such, we can attempt  to find a more efficient representation of the $\widehat{\tau}_{G_l}^{(t)}$ projectors as a process matrix in the bases arising from the local commutants. In turn, such representation will map $\widehat{\tau}^{(t)}$ to a sequence of $L$ such process matrices arranged according to the topology of $\TC$ which we can then interpret as a circuit's TN. Then, if we represent $|\rho^{\otimes t}\rangle \rangle$, and $|O^{\otimes t}\rangle \rangle$ as MPSs in the aforementioned commutants' bases, the expectation value $\mathbb{E}_U[(\ell_U(\rho,O))^t]$ in Eq.~\eqref{eq:momenti-inner-prod} can be computed via standard TNs techniques. As previously mentioned, and as exemplified in our numerics section, having access to the MPS representation of $\widehat{\tau}^{(t)}|O^{\otimes t}\rangle \rangle$ opens up a whole new dimension of analysis, given that we can study its the entropic properties  as it evolved through the TN gates. Notably, such analysis is readily available from our TN approach, but is not  accessible via MC sampling techniques, thus further differentiating our work from previous literature.  

To finish, let us highlight the fact that many of the theoretical and numerical results presented below will focus on the study of the  bond dimension of  $|\rho^{\otimes t}\rangle \rangle$ or $|O^{\otimes t}\rangle \rangle$ as it evolves through the TN. Such analysis will serve several purposes. On the one hand, we recall that the bond dimension is an indicator of the amount of correlations (entanglement) present in the MPS. As such, this parameter controls the ``efficiency'' of the TN contraction algorithms, making it a relevant quantity to report and allowing us to better understand the scaling of our TN simulation techniques. On the other hand, we will show that the study of the bond dimension itself will serve to study the dynamical process of the random circuit, giving operational meaning to this quantity.

\begin{figure}[t]
    \centering
    \includegraphics[width=.9\columnwidth]{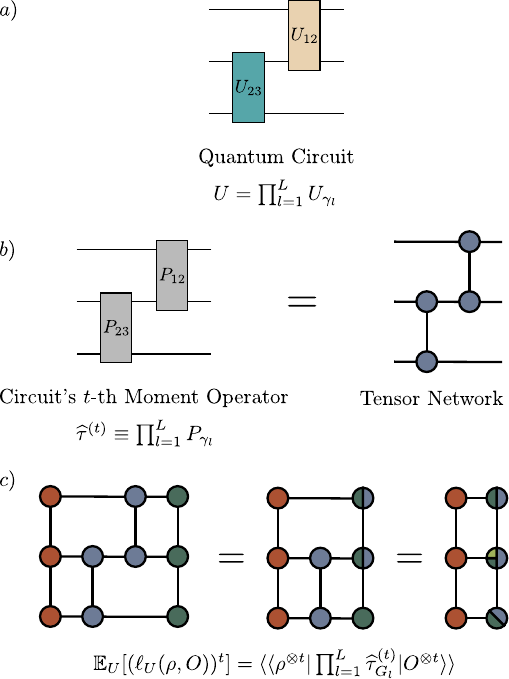}
    \caption{a) We consider a quantum circuit composed of $L$ local  random gates arranged according to some topology $\TC$, and acting on an $n$-qudit system. b) The TN for the ensuing operator $\widehat{\tau}^{(t)}$ will have $n$ legs and will be composed of $L$ gates arranged with the same topology $\TC$. c) By expressing the vectorized operators $|\rho^{\otimes t}\rangle\rangle$ and  $|O^{\otimes t}\rangle\rangle$ as MPSs, we can evaluate the moment by contracting these MPSs with the TN of $\widehat{\tau}^{(t)}$.   }
    \label{fig:tor-model}
\end{figure}

\subsection{Toy model example}\label{sec:toy_model}

To illustrate the previous idea of tackling the problem via TNs, consider an $n=3$ qubit system ($d=2$) where the random quantum circuit is composed of two ($L=2$) two-qubit gates ($k_l=2$ for $l=1,2$) following the topology $\TC=\{(1,2),(2,3)\}$ (see Fig.~\ref{fig:tor-model}(a)).  That is
\begin{equation}\label{eq:unitary-toy}
    U=\prod_{l=1}^2  U_{\gamma_l}=U_{(2,3)}U_{(1,2)}\,.
\end{equation}
Then, let us assume that both local unitaries are sampled i.i.d. from the fundamental representation of $U(4)$. Moreover, we will be interested in evaluating the second moment $\mathbb{E}_U[(\ell_U(\rho,O))^2]$. To finish, let us denote the Hilbert space over which $U$ acts as $\HC=\HC_1\otimes\HC_2$, where $\HC_j$ is the Hilbert space of the $j$-th qubit.

It is well known that
\begin{equation}
    {\rm comm}^{(2)}(G)={\rm span}_{\mathbb{C}}\{\id\otimes \id,{\rm SWAP}\}\,,
\end{equation}
where $\id$ denotes the identity matrix acting on $\HC$, while ${{\rm SWAP}}$ is the operator that exchanges the two copies of the (two-qubit) Hilbert space~\cite{ragone2022representation,larocca2022group,mele2023introduction}. Since the  operators in this basis are non-normalized and non-orthogonal, we find it convenient to instead define the orthonormal basis (under the Hilbert-Schmidt inner product) 
\begin{equation}
   \textrm{comm}^{(2)}(G)={\rm span}_{\mathbb{C}} \left\{\frac{\id\otimes\id}{4},\frac{{\rm S}\otimes\id+\id\otimes {\rm S}+{\rm S}\otimes {\rm S}}{4\sqrt{15}}\right\}\nonumber\,.
\end{equation}
\normalsize
Here, we have used the fact that   we can always express ${\rm SWAP}={\rm SWAP}_1\otimes {\rm SWAP}_2$, with ${\rm SWAP}_j$ the operator that swaps the $j$-th qubits, as well as the fact that \begin{align}
    {\rm SWAP}=\frac{1}{2}\left(\id+{\rm S}\right)\,,\nonumber
\end{align}
with ${\rm S}=X\otimes X+Y\otimes Y+Z\otimes Z$ and $X,Y$ and $Z$ the Pauli operators. Here  we implicitly reordered the order of the Hilbert spaces from $\HC\otimes\HC$ to $\HC_1^{\otimes 2}\otimes\HC_2^{\otimes 2}$, so that ${\rm SWAP}_1$ (${\rm SWAP}_2$) acts on the two copies of the first (second) qubit's Hilbert space.

With the previous, we can express the vectorized orthonormal basis of $\textrm{comm}^{(2)}(G)$ as
\begin{equation}
    \left\{\frac{|\id\rangle\rangle|\id\rangle\rangle}{4},\frac{|{\rm S}\rangle\rangle|\id\rangle\rangle+|\id\rangle\rangle|{\rm S}\rangle\rangle+|{\rm S}\rangle\rangle|{\rm S}\rangle\rangle}{4\sqrt{15}}\right\}\,.
\end{equation} 
In this notation, one can verify that the following equalities hold
\begin{align}
\widehat{\tau}_{U(4)}^{(2)}|\id\rangle\rangle|\id\rangle\rangle&=|\id\rangle\rangle|\id\rangle\rangle\nonumber\,,\\
\widehat{\tau}_{U(4)}^{(2)}|{\rm S}\rangle\rangle|\id\rangle\rangle&=\frac{1}{5}(|{\rm S}\rangle\rangle|\id\rangle\rangle+|\id\rangle\rangle|{\rm S}\rangle\rangle+|{\rm S}\rangle\rangle|{\rm S}\rangle\rangle)\,,\nonumber\\
\widehat{\tau}_{U(4)}^{(2)}|\id\rangle\rangle|{\rm S}\rangle\rangle&=\frac{1}{5}(|{\rm S}\rangle\rangle|\id\rangle\rangle+|\id\rangle\rangle|{\rm S}\rangle\rangle+|{\rm S}\rangle\rangle|{\rm S}\rangle\rangle)\,,\nonumber\\
\widehat{\tau}_{U(4)}^{(2)}|{\rm S}\rangle\rangle|{\rm S}\rangle\rangle&=\frac{3}{5}(|{\rm S}\rangle\rangle|\id\rangle\rangle+|\id\rangle\rangle|{\rm S}\rangle\rangle+|{\rm S}\rangle\rangle|{\rm S}\rangle\rangle)\,,\nonumber
\end{align}
meaning that we can represent the effective action of $\widehat{\tau}_{U(4)}^{(2)}$ as the $4\times 4$ matrix ($2\times 2\times 2\times 2$ tensor) 
\begin{equation}\label{eq:matrix-P}
    \widehat{\tau}_{U(4)}^{(2)}\equiv P=\begin{pmatrix}
        1 & 0 & 0 & 0\\
        0 & 1/5 & 1/5 & 3/5 \\
        0 & 1/5 & 1/5 & 3/5 \\
        0 & 1/5 & 1/5 & 3/5
    \end{pmatrix}\,.
\end{equation}
Importantly, $P$  acts on two, two-dimensional, vector spaces with basis $\{|\id\rangle\rangle,|{\rm S}\rangle\rangle \}$.

From the previous, and as shown in Fig.~\ref{fig:tor-model}(b), the product of these moment operators can be reduced to a $8 \times 8$ matrix (a $2 \times 2 \times 2 \times 2 \times 2 \times 2$ tensor) 
\begin{equation}\label{eq:tensor-toy}
\widehat{\tau}^{(2)}\equiv P_{(2,3)}P_{(1,2)}\,,
\end{equation}
where $P_{(2,3)}$ indicates that the $P$ gate of Eq.~\eqref{eq:matrix-P} acts on the second and third two-dimensional legs. If we compare Eqs.~\eqref{eq:unitary-toy} with~\eqref{eq:tensor-toy}, this example shows that the circuit's moment operator can be expressed as a 
TN composed of $2 \times 2 \times 2 \times 2$ tensors. 
Then, the expectation value $
\mathbb{E}_U[(\ell_U(\rho,O))^2]$ in Eq.~\eqref{eq:momenti-inner-prod} can be estimated by  representing  $|\rho^{\otimes 2}\rangle\rangle$ and $|O^{\otimes 2}\rangle\rangle$ as an MPS in the basis $\{|\id\rangle\rangle,|{\rm S}\rangle\rangle \}^{\otimes 3}$, and evaluating the inner product of these smaller operators (see Fig.~\ref{fig:tor-model}(c)).  For instance if $O$ is a Pauli operator on the first qubit, then 
\begin{equation}\label{eq:O-toy-mode-mps}
    |O^{\otimes 2}\rangle\rangle\equiv \frac{1}{3}|{\rm S}\rangle\rangle|\id\rangle\rangle|\id\rangle\rangle\,,
\end{equation}
which is an MPS of bond dimension $\chi=1$.
Here, we have exploited the fact that from the point of view of twirling over $U(4)$, every Pauli operator is the same, i.e., ${\tau}^{(2)}_{U(4)}(X_1^{\otimes 2}) = {\tau}^{(2)}_{U(4)}(Y_1^{\otimes 2}) = {\tau}^{(2)}_{U(4)}(Z_1^{\otimes 2}) = \frac{1}{3} {\tau}^{(2)}_{U(4)}({\rm S})$.

\subsection{General formalism}

The previous example illustrates how $
\mathbb{E}_U[(\ell_U(\rho,O))^t]=\langle\langle \rho^{\otimes t}|\widehat{\tau}^{(t)}|O^{\otimes t}\rangle \rangle$ can be computed by representing $\widehat{\tau}^{(t)}$ as an $n$-legged TN and $|\rho^{\otimes t}\rangle \rangle$ and $|O^{\otimes t}\rangle \rangle$ as MPSs.  In fact, just like in the previous toy model, given a  unitary with random local gates of the form $U=\prod_{l=1}^L  U_{\gamma_l}$, the TN for $\widehat{\tau}^{(t)}$ will have $n$ legs and will be composed of $L$ gates --which we will dub ``$P$ gates''-- arranged according to the topology $\TC$. That is
\begin{equation}
    \widehat{\tau}^{(t)}=\prod_{l=1}^LP_{\gamma_l}\,.
\end{equation}
Each $P_{\gamma_l}$ gate will have $k_l$ legs, and its input and output dimensions will depend on who the local group $G_l$ is, but also on the $P$ gates that precede and follow it (as it needs to capture how the local commutants get projected into each other). In the simpler case  when $k_l=k$ $\forall l$ and when all the local groups are the same, the $P$ gates will have the same dimensions.  

Note that the previous does not imply that we can efficiently contract the tensors, as this will ultimately depend on the local tensor dimensions, the circuit topology, and the entanglement in the MPSs arising from the initial state and measurement operator. In what follows we present a few general considerations that will allow us to better understand when the TN simulation might be efficient.

We begin by analyzing the dimensions of the $P$ gates for the case when the local unitaries are sampled from several Lie groups of interest. In particular, we will focus on the case when $G$ is the unitary, orthogonal, and symplectic groups, and also from a free-fermionic group. We study the first three aforementioned groups as these are the basic groups appearing from the classification of (compact) Lie groups~\cite{fulton1991representation}, while the last one is of interest to the quantum information community due to the fact that it leads to classically simulable circuits~\cite{jozsa1998universal}, as well as trainable quantum neural networks~\cite{diaz2023showcasing}. 

To begin, we can show that the following result holds. 

\begin{theorem}\label{theo:fundamental}
    If the $k$-local unitaries are sampled i.i.d. from the  fundamental representation of the Lie group $G\in\{U(d^k),O(d^k),Sp(d^k/2)\}$, where $U(\cdot)$ stands for unitary, $O(\cdot)$ for orthogonal and $Sp(\cdot)$ for symplectic (assuming $d$ is even), then the $P$ gate for each $\widehat{\tau}_{G_l}^{(t)}$  will be a square matrix of dimension up to $(t!)^2\times (t!)^2 $ for $G=U(d^k)$, and of dimension up to $\left(\frac{(2t!)}{2^tt!}\right)^2\times \left(\frac{(2t!)}{2^tt!}\right)^2 $, for $G=O(d^k),Sp(d^k/2)$. 
\end{theorem}
The proof of this theorem, as well as that of our other main theoretical results, can be found in the appendices.

For the special case of a circuit composed of two-qubit gates ($d=2$ and $k=2$), Theorem~\ref{theo:fundamental} implies that the first moment operator will be trivial for any $G\in\{U(d^k),O(d^k),Sp(d^k/2)\}$, as the $P$-gates for each $\widehat{\tau}_{G_l}^{(1)}$ are one dimensional projectors onto the vectorized identity (which follows from the fact that the local groups are irreducible). Then, the second moment operator $\widehat{\tau}^{(2)}$  will be composed of $4\times 4$   square gates (or $2\times2\times2\times 2$ tensors) for  $G=U(4)$ given by Eq.~\eqref{eq:matrix-P}, and of dimension $9\times 9$ (or a $3\times3\times3\times 3$ tensor)  for $G=O(4),Sp(2)$. We note that  the dimension bounds given by Theorem~\ref{theo:fundamental} will be useful for small $t$ or large qudit dimension $d$, but these can also be extremely loose. In this case, one can tighten them by using representation theoretical tools. For instance, if $d=2$ and $G$ is the unitary group, one should replace the upper bound of $(t!)^2\times (t!)^2 $ by $(C_t)^2\times (C_t)^2$, where $C_t$ are the Catalan numbers\footnote{This follows from the fact that the order $t!$ of the symmetric group $S_t$ can be much larger than the operator's space dimension, leading to an over-complete basis. Here, one needs to instead compute how many trivial irreducible representations appear in the operator space decomposition under the adjoint action of the $t$-th fold representation. }. 

It is important to note that while the dimension of the tensor representation of  $\widehat{\tau}_{G_l}^{(t)}$ matches $(\dim({\rm comm}^{(t)}(G)))^2\times (\dim({\rm comm}^{(t)}(G)))^2$ for these groups\footnote{Here we used the Schur-Weyl duality between the $t$-fold representation of the unitary group and the symmetric group $S_t$, as well as that for the $t$-th fold representation of the orthogonal (or symplectic) group and the Brauer algebra $B_{2t}$. See for instance~\cite{goodman2009symmetry,garcia2023deep}.}, this is not a generic feature. In fact, the dimension of the $\widehat{\tau}_{G_l}^{(t)}$ tensor can depend on who $\rho$ and $O$ are. For instance, if the unitary's two-qubit gates are sampled from the free-fermionic reducible representation of $G=SO(4)\simeq SU(2)\oplus SU(2)$~\cite{kokcu2022algebraic,wiersema2023classification,diaz2023showcasing,oszmaniec2022fermion,pozsgay2024quantum}, then we have that $\dim({\rm comm}^{(2)}(G))=10$~\cite{diaz2023showcasing}, which would imply that $\widehat{\tau}_{G_l}^{(2)}$ are matrices of size $100\times 100$ if one were to extrapolate the previous realization. However, we find that the following theorem holds.   
\begin{theorem}\label{theo:-free-fermion}
    If the $k$-local unitaries are sampled i.i.d. from the free-fermionic reducible representation of the Lie group $G=SO(4)\simeq SU(2)\oplus SU(2)$ then the $P$ gate for each $\widehat{\tau}_{G_l}^{(t)}$  will be a square matrix of dimension  $9\times 9$ (or a $3\times3\times3\times 3$ tensor) if $\rho$ has fixed parity, and of size $36\times 36$ (or a $6\times6\times6\times 6$ tensor) if $\rho$ is a generic state. 
\end{theorem}

Next, let us analyze the bond dimension of the MPSs arising from $\rho$ and $O$.  
If $\rho$ is a product state and if $O$ can be expressed as a tensor product of Hermitian operators (e.g., a projector onto a computational basis state, or a Pauli operator), then $|\rho^{\otimes t}\rangle \rangle$ and $|O^{\otimes t}\rangle \rangle$ can be represented as MPSs with bond dimension $\chi=1$. From here one can wonder how large the bond dimension of $\widehat{\tau}^{(t)}|O^{\otimes t}\rangle \rangle$ or $\widehat{\tau}^{(t)}|\rho^{\otimes t}\rangle \rangle$ will be as a function of the number of gates $L$ in the circuit, and concomitantly of the $P$-gates  in the TN. While the bond dimension for small $L$ can vary widely depending on the local groups and topology (see the numerics section) we can make statements about the large $L$ limit. 

In particular, let us denote as $G_U$ the Lie group to which the distribution of $U$ converges to for large $L$. We can obtain $G_U$ as follows. First, let $\GC_l=\{iH_{lm}\}_{m=1}^{\dim(\mathfrak{g}_l)}$ be a basis for the Lie algebra $\mathfrak{g}_l$ associated to the Lie group $G_l$ (that is $G_l=e^{\mathfrak{g}_l}$). Then, we need to compute the Lie algebra~\cite{zeier2011symmetry} 
\begin{equation}
    \mathfrak{g}_U=\left\langle \bigcup_{l=1}^L \GC_l \right \rangle_{{\rm Lie}}\,,
\end{equation}
where $\langle \cdot \rangle_{{\rm Lie}} $ denotes the Lie closure, i.e., the vector space obtained by nested commutators of all the operators in the bases $\GC_l$. Such Lie algebra is important as  $U\in G_U=e^{\mathfrak{g}_U}$ for any set of randomly sampled local gates. Moreover, as the number of gates increases, one can generally expect that $\mathbb{E}_U=\mathbb{E}_{G_L}\cdots \mathbb{E}_{G_1} \simeq \mathbb{E}_{G_U}$.  The number of gates needed for the circuit to become a $t$-design over $G_U$ can be estimated via the spectral gap of $\widehat{\tau}^{(t)}$, i.e., how large its first not-equal-to-one  eigenvalue is~\cite{larocca2021diagnosing,ragone2023unified,harrow2009random,brandao2016local,harrow2018approximate,Haferkamp2022randomquantum,hunter2019unitary,haah2024efficient,chen2024efficient}. 

The previous shows that if $L$ is large enough so that the distribution of $U$ becomes a $t$-design over $G_U$, then  $\widehat{\tau}^{(t)}$ becomes a projector onto  ${\rm comm}^{(t)}(G_U)$. Then, how large the bond dimension of $\widehat{\tau}^{(t)}|O^{\otimes t}\rangle \rangle$ or $\widehat{\tau}^{(t)}|\rho^{\otimes t}\rangle \rangle$ is will solely depend on what's the bond dimension of a linear combination of the vectorized operators in the basis of this commutant. Thus, we find that the following proposition holds.
\begin{proposition}\label{prop:bond-dimension}
    Let $U$ be deep enough so that it forms a $2$-design over $G_U$. The  bond dimension of the MPS $\widehat{\tau}^{(t)}|A^{\otimes t}\rangle \rangle$ for any $A\in\BC(\HC)$ is up to $\chi=t!$ for $G_U=U(d^n)$ and up to $\chi=\frac{(2t!)}{2^tt!}$ for $G=O(d^n),Sp(d^n/2)$. 
\end{proposition}

When $t=2$,  Proposition~\ref{prop:bond-dimension} implies that irrespective of the number of qudits $n$, or the local dimension $d$, The maximum bond dimension of the MPS $\widehat{\tau}^{(t)}|A^{\otimes t}\rangle \rangle$ for any $A\in\BC(\HC)$ is $\chi=2$ for $G_U=U(d^n)$ and $\chi=3$ for $G=O(d^n),Sp(d^n/2)$. 

It is important to note that the results in the previous proposition follow from the fact that the elements in the commutants of $U(d^n),G=O(d^n),Sp(d^n/2)$ can be naturally expressed as rank-one projectors onto MPSs of bond dimension $\chi=1$ for the Hilbert space $\HC_1^{\otimes t}\otimes \HC_2^{\otimes t}\cdots \otimes \HC_n^{\otimes t}$, where $\HC_j$ denotes the Hilbert space of the $j$-th qudit.  Such result will not generally hold, especially for the case when $G_U$ is not an irreducible fundamental representation of a Lie group. In the more general scenario, one needs to find how the vectorized elements of ${\rm comm}^{(t)}(G_U)$ will decompose in terms of operators acting on $\HC_1^{\otimes t}\otimes \HC_2^{\otimes t}\cdots \otimes \HC_n^{\otimes t}$. Such analysis will be case by case dependent, but we can provide some guidelines for some special cases.   

In particular, let us study the situation when  $i\rho$ or $iO$ belonging to $\mathfrak{g}_U$, and when $t=2$. We recall that we can always express $\mathfrak{g}_U$ in its reductive decomposition, i.e., as a direct sum of commuting ideals 
\begin{equation}\label{eq:ideal-decomp}
    \mathfrak{g}_U=\mathfrak{g}_1\oplus  \cdots\oplus \mathfrak{g}_{\zeta-1}\oplus\mathfrak{g}_{\zeta}\,,
\end{equation}
where $\mathfrak{g}_{j}$ are simple Lie algebras for $j=1,\ldots,\zeta-1$, and with $\mathfrak{g}_{\zeta}$ being the center of $\mathfrak{g}_U$, and therefore abelian. Here, the only elements of the $t$-th fold commutant that will have non-zero overlap with $\rho^{\otimes 2}$ (or $O^{\otimes 2}$)  are the Casimir operators $\{C_j\}_{j=1}^\zeta$~\cite{ragone2023unified}, where
\begin{equation}
    C_j=\sum_{\mu=1}^{\dim(\mathfrak{g}_j)} B_\mu \otimes B_\mu\,,
\end{equation}
and with $\{B_\mu\}_{\mu=1}^{\dim(\mathfrak{g}_j)}$ being an orthonormal basis for $i\mathfrak{g}_j$. For instance, we can use this insight to prove that 
\begin{proposition}\label{prop:bond-dimension-ideals}
    Let $U$ be deep enough so that it forms a $2$-design over the free-fermionic representation of $G_U=SO(2n)$, with associated Lie algebra $\mathfrak{g}_u=\mathfrak{so}(2n)$. Given some Pauli operator $A\in\BC(\HC)$ such that $iA\in\mathfrak{g}_U$, then the maximum bond dimension of the MPS $\widehat{\tau}^{(2)}|A^{\otimes 2}\rangle \rangle$  is $\chi=3$. 
\end{proposition}

Propositions~\ref{prop:bond-dimension} and ~\ref{prop:bond-dimension-ideals}  show an interesting property of our proposed TN approach. In many cases,  for shallow circuits with small $L$, the bond dimension of the MPS is generally small. This follows from the fact that the MPS  representation of $|\rho^{\otimes t}\rangle \rangle$ and $|O^{\otimes t}\rangle \rangle$ can be such that $\chi=1$, and since  the gates in the circuit are local, they cannot generate too much entanglement. Then, for deep circuits with very large $L$  the bond dimension of $\widehat{\tau}^{(t)}|O^{\otimes t}\rangle \rangle$ (or $\widehat{\tau}^{(t)}|\rho^{\otimes t}\rangle \rangle$) is small again.

\section{Dealing with unnormalized tensor network contractions}

The general formalism derived in the previous section  allows us to compute $t$-th moment of an expectation value by building a tensor network. Here, one replaces the gates appearing in $U$ with $P$ tensors arising from the  $\widehat{\tau}_{G_{l}}^{(t)}$,  and replaces local qudits Hilbert spaces $\HC_l$ with legs whose dimension depends on the  local gate's commutants.  For the sake of simplicity, we will refer to the resulting TN as the $P$-network, or $P$-net for short.

While this construction allows us to drastically reduce the  dimension of the problem (see Theorems~\ref{theo:fundamental} and~\ref{theo:-free-fermion}), care must be taken when dealing with the ensuing tensors, as neither the MPS encoding the (vectorized) $t$-th tensor product of the measurement operator $|O^{\otimes t}\rangle \rangle$ nor the one encoding the quantum state $|\rho^{\otimes t}\rangle \rangle$ are in principle normalized states. Similarly, the $P$-gates in the $P$-net need not be unitary (see for instance the $P$-gate in Eq.~\eqref{eq:matrix-P}). 
Hence, when evolving $|O^{\otimes t}\rangle \rangle$ (or $|\rho^{\otimes t}\rangle \rangle$) through the $P$-net, one could very quickly start facing very small or very large numbers, leading to loss of numerical precision or overflow errors. In order to prevent such issues we propose the following recipe to contract the TN:   
\begin{itemize}
    \item We associate a vector $\mathbf{n}_O$ to $|O^{\otimes t}\rangle \rangle$, whose length is equal to the size $n$ of the system, and we fill it with ones. This vector will hold all the normalization factors of the measurement operator's MPS. After the application of $P$-gates acting in parallel, or after a given (fixed) number of gates in case of an unstructured topology, we perform a Singular Values Decomposition (SVD) sweep through $|O^{\otimes t}\rangle \rangle$, normalizing the singular values at each pair of sites. Namely, calling $A^{O}_j$ the $j$-th tensor of the MPS $|O^{\otimes t}\rangle \rangle$, we contract adjacent tensors $A^{O}_j, A^{O}_{j+1}$, obtaining a two-sites tensor $\Tilde{A}^{O}_{j,j+1}$ which we then decompose using SVD as, dropping super and subscripts for the sake of notation, $\Tilde{A}=W_1\cdot S\cdot W_2$, with $W_1$ and $W_2$ unitary tensors and $S$ a diagonal matrix containing the singular values of $\Tilde{A}$. We normalize $S\leftarrow S/\norm{S}$, distributing the weight equally between $W_1$ and $W_2$ as $W_1\leftarrow W_1\sqrt{S}$ and $W_2\leftarrow \sqrt{S}W_2$ and update $A^{O}_j \gets W_1$ and $A^{O}_{j+1}\gets W_2$. In the process we also update the normalization factors as $(\mathbf{n}_O)_j\gets (\mathbf{n}_O)_j \norm{S},(\mathbf{n}_O)_{j+1}\gets (\mathbf{n}_O)_{j+1} \norm{S}$. This procedure allows us to keep the norms of the tensors in $\psi$ well-behaved.
    \item We associate an analogous vector $\mathbf{n}_\rho$ to  $|\rho^{\otimes t}\rangle \rangle$. We then normalize the tensors $A^{\rho}_j$ to $\sqrt{\chi_j}$, $\chi_j$ being the bond dimensions of the link between sites $j$ and $j+1$, and we save the normalization factors in $\mathbf{n}_\rho$ as $(\mathbf{n}_\rho)_j = \norm{A^{\rho}_j}/\sqrt{\chi_j}$. Notice that the last tensor will be normalized to one, as the non-existing link at the right of site $n$ can be treated as having dimension $\chi_n=1$. The intuition behind this is that if $|\rho^{\otimes t}\rangle \rangle$ were to be a normalized quantum state's MPS, in the left canonical form it would be composed of unitary tensors, whose norms are $\sqrt{\chi}$.
    \item Lastly, we compute the inner product $\langle\langle \rho^{\otimes t} |\widehat{\tau}^{(t)}|O^{\otimes t}\rangle \rangle$ by initializing yet another vector $\mathbf{n}_t$ and filling it with the norms of the sequential contractions of pairs of tensors $A^{O}_j, A^{\rho}_j$. Here, we start with the first tensors ($j=1$), contract them, normalize the result, save the norm in $(\mathbf{n}_t)_1$, then proceed to contract this normalized tensor with the second tensors and so on. 
\end{itemize}
Once all the tensor contractions have been performed we are left with three vectors $\mathbf{n}_O$, $\mathbf{n}_\rho$, $\mathbf{n}_t$. The overlap $\langle\langle \rho^{\otimes t} |\widehat{\tau}^{(t)}|O^{\otimes t}\rangle \rangle$ is then obtained as the multiplication of all the entries of said vectors.
Following this recipe, the evolution of $|O^{\otimes t}\rangle \rangle$ through the $P$-net is kept under control and numerically stable. Moreover, we empirically observe that the components of the three vectors $\mathbf{n}_O$, $\mathbf{n}_\rho$, and $\mathbf{n}_t$ are $O(1)$, leading to stable results for the overlaps considered in this paper.

\section{Exponential advantage over MC simulations}

In this section we show that our TN formalism allows us to compute certain quantities  exponentially faster than a MC simulation. In particular, we will be interested in Heisenberg evolving the measurement operator, and studying how it decomposes into Paulis according to their bodyness. That is, we are interested in analyzing whether $U^\dagger O U$ will have support mostly on local operators, or on global ones. In particular, such analysis is fundamental to the barren plateau analysis, and to its connection to classical simulability~\cite{cerezo2023does}.

For this purpose, we will compute the inner product between $U^\dagger O U$ and  all possible Paulis $P_j$ in a basis of $\BC(\mathbb{C}^{2^n})$ of bodyness $|P_j|=k$ (that is, $P_j$ acts non trivially on $k$ qubits).  These quantities of interest will henceforth be called $k$-purities, denoted as $p^{(k)}_{O}$, and take the form
\begin{equation}\label{eq:k-puritites}
    p^{(k)}_{O} = \frac{1}{4^n} \mathbb{E}_{U} \left[ \sum_{P_j\,:\,|P_j|=k} \Tr [P_j U^\dagger O U]^2\right] \, .
\end{equation}
It is not hard to see that since $O$ is itself a Pauli, then $\sum_{k=1}^n p^{(k)}_{O}=1$, and since by definition $p^{(k)}_{O}\geq 0$ for all $k$ , then the $k$-purities form a  probability distribution. Next, let us note that there will exist cases for which $p^{(k)}_{O}$ is an exponentially small quantity. For instance, if the circuit $U$ is deep enough so that it forms a $2$-design, then one can readily use Eq.~\eqref{eq:twirl-vectorized} to find that 
\begin{equation}\label{eq:purity-Haar}
p^{(k)}_{O}=\frac{3^k\binom{n}{k}}{4^n-1}\qquad \forall k\geq 1\,.
\end{equation}
As explicitly discussed in the appendix, our TN formalism is able to exactly compute the value of $p_O^{(k)}$  by means of the following calculation: \begin{equation}\tag{22}
    p_O^{(k)} = \langle\langle \phi_k \vert \widehat{\tau}^{(2)} \vert O^{\otimes 2}\rangle \rangle \,,\\
\end{equation}
where $\langle\langle \phi_k \vert$ is a projector in the form of an MPS onto the subspace of all Paulis with a given bodyness $k$, $\widehat{\tau}^{(2)}$ is the $P$-net, and $\vert O^{\otimes 2}\rangle \rangle$ is the MPS encoding the vectorized two copies of the measurement operator $O$ projected into the local commutants basis. 
This TN algorithm can be divided into three distinct steps.
First, the application of the $P$-net layers to $\vert O^{\otimes 2}\rangle \rangle$. 
Let us dub $\chi$ the maximum bond dimension of $\vert O^{\otimes 2}\rangle \rangle$ as it evolves through the layers of the network and find the computational complexity of carrying out this step \cite{orus2014practical}. To apply each $P$-gate we first express it as an MPO of bond dimension $d_{G,2}$ equal to the rank of the gate itself, which corresponds to the size of the second order commutant of the representation of the group $G$ at hand, i.e. $d_{G,2}=2$ for the unitary group and $d_{G,2}=3$ for the orthogonal one. Then we apply this MPO to the sites targeted by the $P$-gate, and all those in-between if the gate is more than 2-local, incurring in a worst case computational cost of $\OC(\gamma\chi^3 d_{G,2}^4)$, where we denoted the locality of the $P$-gate with $\gamma$. 
After applying a full layer of non-intersecting $P$-gates, we perform the second step of our algorithm, namely a singular value decomposition (SVD) lossless compression of the evolved MPS to optimize its bond dimension $\chi$. To do this, we pairwise contract adjacent tensors in the MPS and then split them again via SVD, without discarding any singular values above machine precision, hence why lossless. This step's complexity scales as $\OC(n d_{G,2}^3\chi^3)$.
Lastly, we project the backpropagated operator onto a given subspace of Paulis with bodyness $k$ by means of the vectorized operator $\langle\langle \phi_k \vert$ (whose maximum bond dimension $\chi'$ is shown to be upper bounded by $\lfloor n/2 \rfloor$ in the appendix). This last step has a cost upper bounded by $\OC(n d_{G,2} \chi^2\chi')$ or $\OC(n d_{G,2}\chi'^2\chi)$, depending on the pair $\chi$ and $\chi'$.

We can hence see that at every step we pay a polynomial price in the MPS bond dimension $\chi$. If we can keep it scaling polynomially in the system size $n$, our algorithm will be itself polynomial. Each layer of non-intersecting gates can at most increase $\chi$ by a factor $d_{G,2}$, hence a circuit composed of $n_L$ such layers will lead to $\chi$ scaling as $\OC(d_{G,2}^{n_L})$. Of the two kinds of architecture that we will consider in the following numerical investigation, one has $n_L\in \OC(\log(n))$, whereas for the second one we find that $\chi\in\OC(1)$ for the unitary group.

With the previous, we can show that our algorithm is capable of achieving an exponential advantage against MC for computing the $k$-purities. In particular, we can readily see that using a MC simulation to obtain an absolute error smaller than $\epsilon$ (with respect to the TN simulation) requires a number of samples scaling as $\Theta(16^n)$. As such, given that every sample requires a computational complexity of $\OC(\poly(n))$, the MC simulation is exponentially expensive. This follows from a standard Chernoff bound~\cite{napp2022quantifying}, which indicates that MC can estimate a quantity with additive error at most $\varepsilon$ and with  probability of failure of at most $\delta$, via a number of samples $n_s$ scaling as $\Theta(\frac{1}{\varepsilon^2}\log(\frac{1}{\delta}))$. Therefore, setting $\delta$ as constant and noting that for $k\in\OC(1)$ Eq.~\eqref{eq:purity-Haar} implies that the purities (and thus $\varepsilon$) will be in $\Theta\left(\frac{1}{4^n}\right)$, leads to $n_s\in\Theta(16^n)$.

\section{Numerical Results}

In this section, we showcase our TN framework by computing the exact $t=2$ moments for two well-known qubit ($d=2$) circuit topologies: a Quantum Convolutional Neural Networks (QCNN)~\cite{cong2019quantum,pesah2020absence} and a one-dimensional Hardware Efficient Ansatz (HEA)~\cite{kandala2017hardware,cerezo2020cost,leone2022practical}, when all the gates composing $U$ on each architecture are sampled i.i.d. first from the fundamental representations of $G=U(4)$ and from $G=O(4)$. The circuit topologies are shown in Fig.~\ref{fig:topologies}. Importantly, as we will see below, when the local gates are sampled from $O(4)$, MC simulations are not available due to a sign problem. All the numerical analyses carried out in the following have been obtained with custom code built on top of the ITensors.jl software library \cite{fishman2022itensor} for the Julia programming language \cite{bezanson2017julia}.

\begin{figure}[t]
    \centering
    \includegraphics[width=.9\columnwidth]{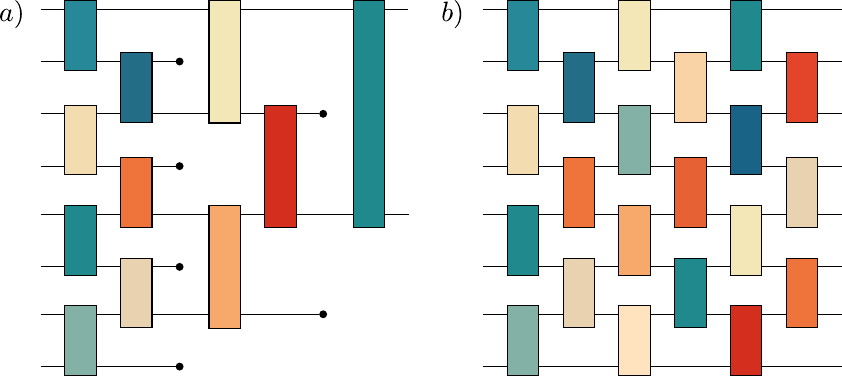}
    \caption{Circuit topologies for the numerical experiments. a) A QCNN is composed of a ``convolutional'' layer where two-qubit gates on act on alternating pairs of qubits, and of ``pooling'' layers, where half of the qubits are traced out. QCNNs naturally have a number of layers $n_L$ in $\OC(\log(n))$, and a number of local gates in $\OC(n)$. b) In each layer of an HEA, gates act on alternating neighboring qubits in a brick-like fashion. Here we show an HEA with $n_L=3$ layers with open boundary conditions. An HEA with $n_L$ layer, has a number of gates in $\OC(n n_L)$.  }
    \label{fig:topologies}
\end{figure}

\subsection{Local gates sampled from $G = U(4)$}\label{sec:U-results}

In this section we will study the second moment for the case when $O$ is a  Pauli operator. In particular, we will evaluate the $k$-purities of Eq.~\eqref{eq:k-puritites}.

As it was previously discussed, the core of our TN calculations rely on choosing an adequate basis for the local commutants, as these will define the gates for the $P$-net. For the case of $U(4)$, we have already derived in Eq.~\eqref{eq:matrix-P} the matrix representation of the $P$-gates for the particular choice of the local basis $\{ |\id \rangle\rangle, |{\rm S}\rangle\rangle\}$.

\subsubsection{Quantum convolutional neural network}
We begin by taking $U$ to be a  QCNNs as introduced in \cite{cong2019quantum}, with the measurement operator chosen to be $O=Z_1$. To truly illustrate the power of our techniques, we consider a system of $n=1264$ qubits. Clearly, a full density matrix simulation of the expectation values would be beyond any plausible supercomputer as it would require over $10^{1632}$ bits to save each amplitude to machine precision\footnote{That is, we would need a device with over sixteen googol bits just for memory; or we would need to store $\sim20$ bits of information in every atom of the visible universe.}.

\begin{figure}
    \centering
    \includegraphics[width=1\columnwidth]{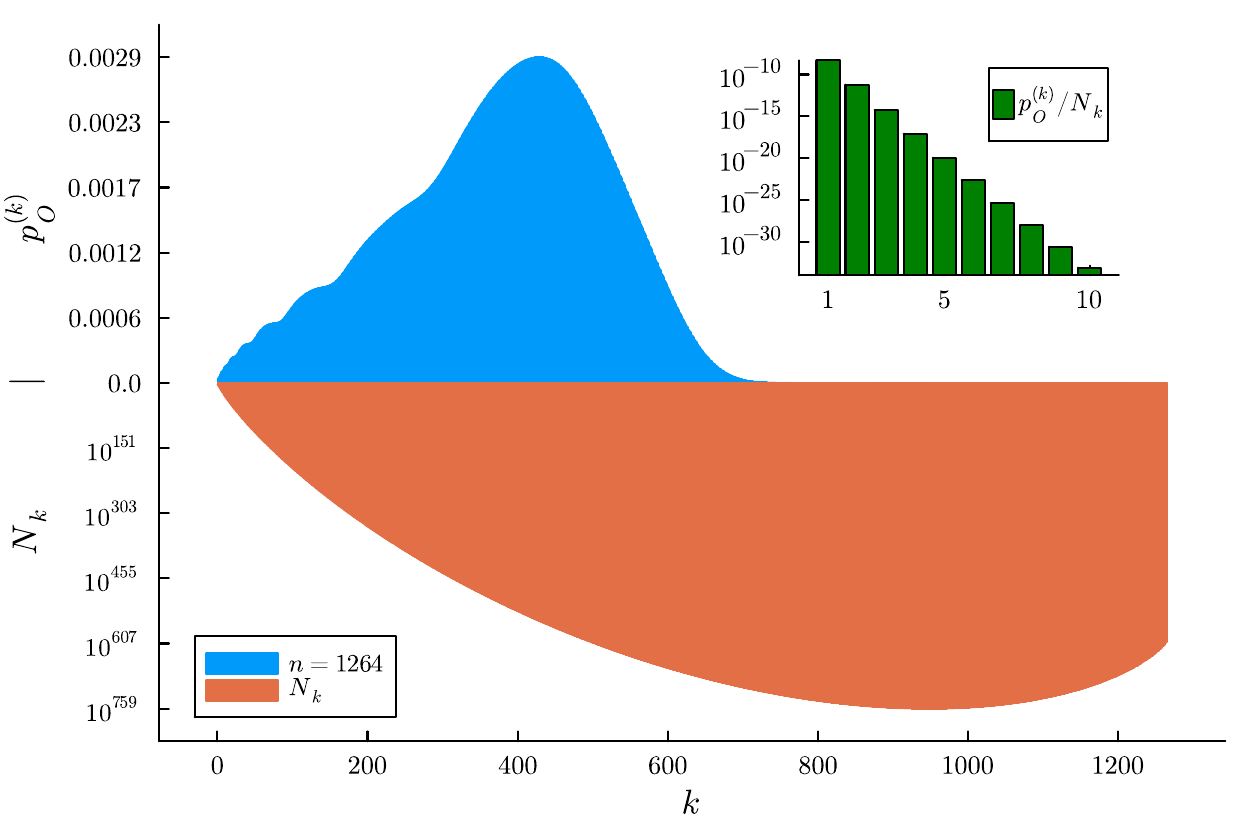}
    \caption{ We show the distribution of $k$-purities of Eq.~\eqref{eq:k-puritites} for a QCNN acting on $n=1264$ qubits (blue). We also present the number of Paulis $N_k=3^k \binom{n}{k}$ acting on $k$-qubits (orange). The inset depicts the quotient $p^{(k)}_{O}/N_k$. }
    \label{fig:huge_qcnn}
\end{figure}

In Fig.~\ref{fig:huge_qcnn}(blue), we show the $k$-purities of the QCNN acting on $n=1264$ qubits. 
Here, we can see that the  vast majority of contributions to the $k$-purities are  concentrated in a region with $k\sim n/3$. Moreover, it is clear that the distribution is skewed towards more local operators, and has essentially no contributions coming from global Paulis acting on all qubits. Of course, care must be taken when interpreting this result as we recall that it shows the cumulative purity for all Paulis of a given bodyness. If we wanted to ask: What is the contribution to the purity coming from  a single Pauli acting on $k$-qubits, we would have to roughly divide $p^{(k)}_{O}$ by the number of Paulis $N_k=3^k \binom{n}{k}$ of a given bodyness. Thus, since there are significantly more Paulis with bodyness $k=n/3$ than Paulis with bodyness $k=1$ (see Fig.~\ref{fig:huge_qcnn}(orange)), we can expect that the contribution of a  Pauli acting on a single-qubit must be much larger than that of a single Pauli acting on $k$ qubits. In fact, we have shown in the inset of  Fig.~\ref{fig:huge_qcnn} the quantity $p^{(k)}_{O}/N_k$, and we can indeed see that the contributions to the purities  decay exponentially with $k$.

\subsubsection{One-dimensional hardware efficient ansatz}

\begin{figure}
    \centering
    \includegraphics[width=1\columnwidth]{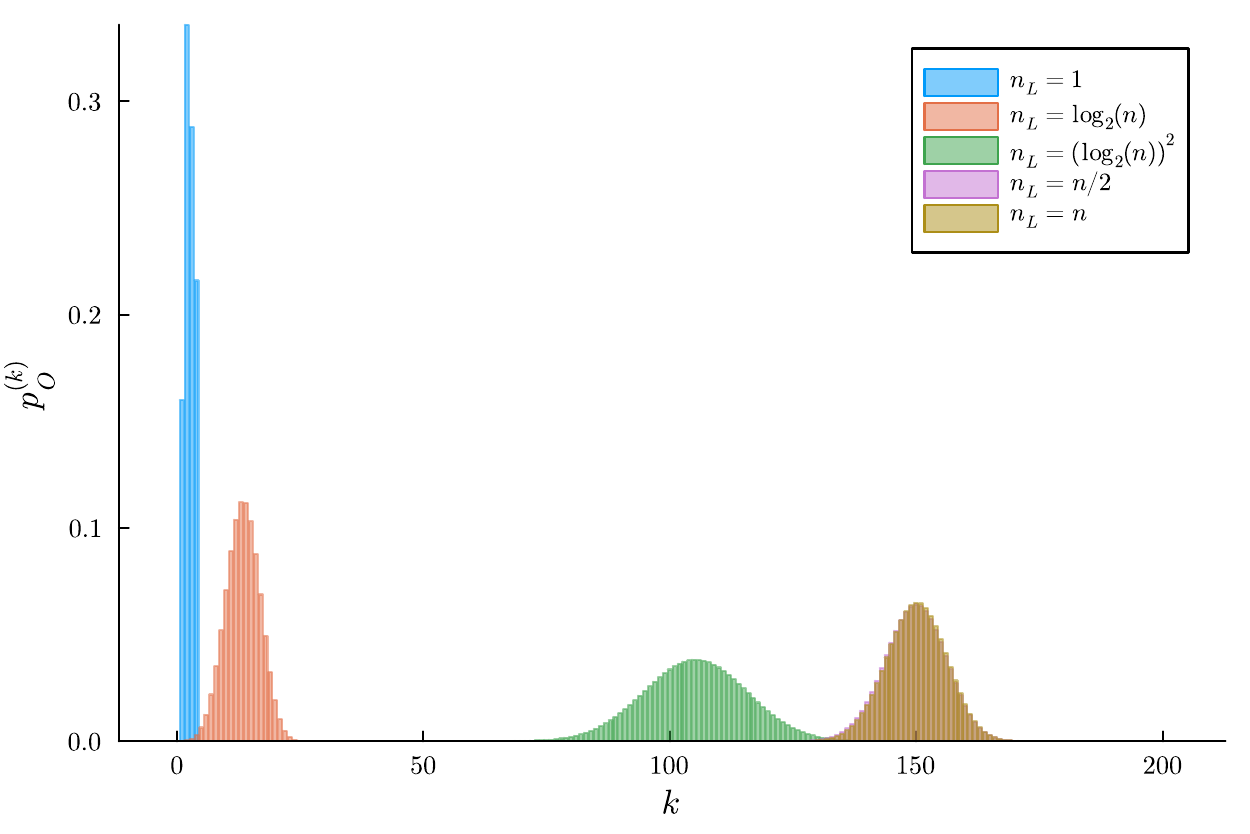}
    \caption{ We show the distribution of $k$-purities of Eq.~\eqref{eq:k-puritites} for a one-dimensional HEA acting on $n=200$ qubits as a function of the number of layers $n_L$. We note that the distributions for $n_L=n/2$ and $n_L=n$ are completely indistinguishable. }
    \label{fig:deep_su4_hea}
\end{figure}

Let us now proceed to study the $k$-purities for an HEA acting on $n=200$ qubits. In this example, the operator measured at the end of the circuit will be taken to be $O=Z_{n/2}$,   Pauli-$z$ operator acting on the middle qubit. Note that unlike the QCNN where the number of layers is fixed, in an HEA, the depth of the circuit is a free parameter. As such, in Fig.~\ref{fig:deep_su4_hea}  we show  how the $k$-purities change as a function of the HEA's number of layers $n_L$. Here, we can see that --as expected due to the circuit's light-cone structure~\cite{cerezo2020cost}-- for shallow circuits $n_L=1$  the Heisenberg evolved operator $U\ad OU$ is mostly local. This is evidenced from the fact that $k$-purities for $n_L=1$ concentrate at $k\sim 2$. Then, as the number of layers increases, we can see that the distribution of $k$-purities shifts and peaks at  higher values of $k$.  Eventually, the distribution converges at an approximate Gaussian distribution with mean in $k\sim 3n/4$ (i.e., the purity distribution of $n_L=n/2$ and $n_L=n$ are completely indistinguishable and match that of Eq.~\eqref{eq:purity-Haar}). This is completely expected as it is known that one-dimensional hardware efficient ansatz with local random gates forms a $2$-design over $SU(2^n)$ for $n_L\in\OC(n)$~\cite{harrow2009random,brandao2016local,harrow2018approximate}. Such convergence is evidenced in our plots, and seen to be reached as early as $n_L=n/2$. 

\subsubsection{Maximum bond dimension scaling}

\begin{figure}
    \centering
    \includegraphics[width=1\columnwidth]{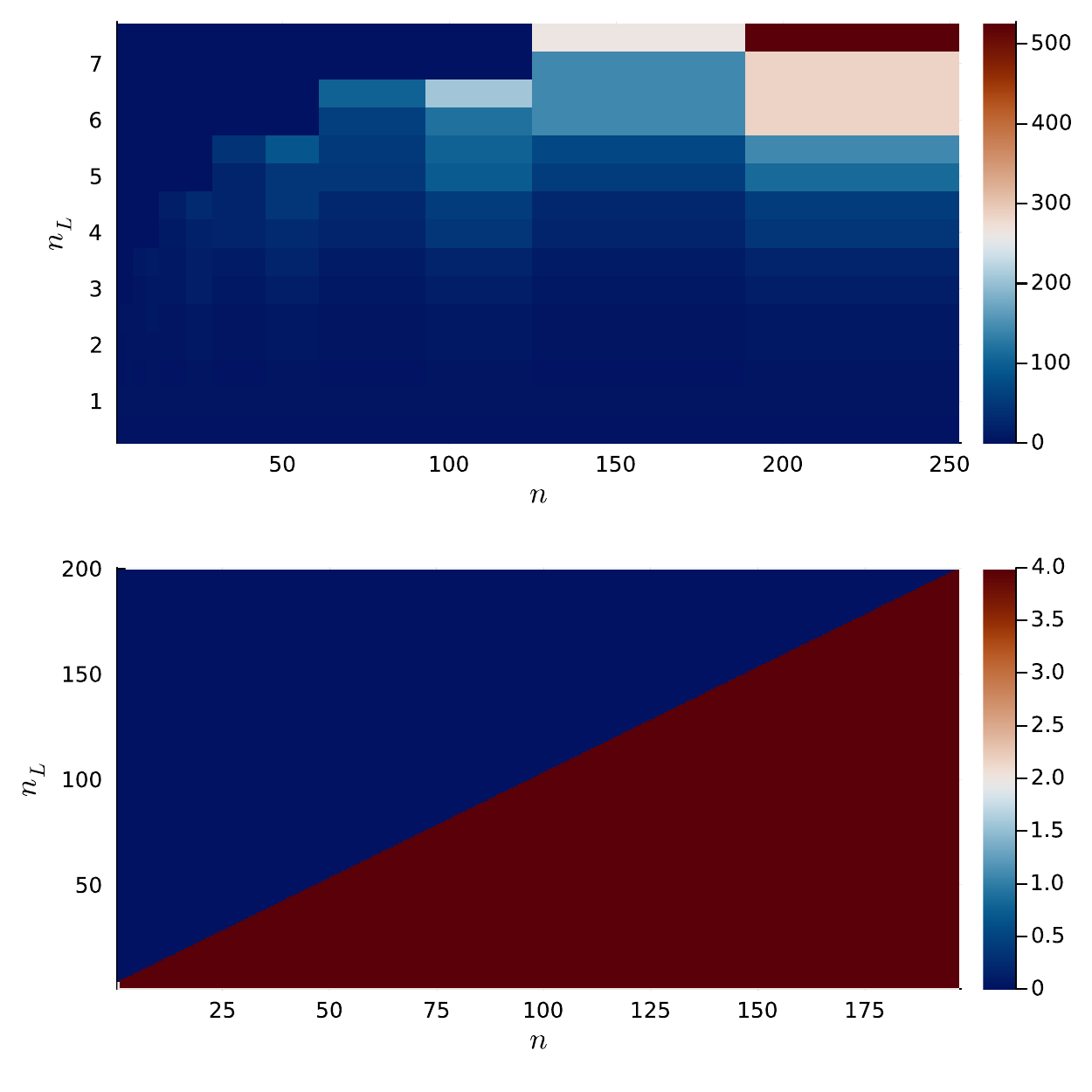}
    \caption{Scaling of the maximum bond dimension $\chi_{\rm max}$ of the MPS $|O^{\otimes t}\rangle \rangle$ for the QCNN circuit topology (top) and the HEA (bottom) as a function of the number of qubits $n$ and the number of layers $n_L$.}
    \label{fig:su4_hea_qcnn_maxchi}
\end{figure}

The ability of our method to deal with  QCNNs and deep HEAs acting on large number of qubits stems from the good behavior of the bond dimension $\chi$ of the MPS representing the evolution of the input measurement operator $|O^{\otimes t}\rangle \rangle$ throughout the $P$-net. In Fig.~\ref{fig:su4_hea_qcnn_maxchi}  we thus plot the maximum value $\chi_{\rm max}$ of the vectorized measurement operator MPS, different number of qubits $n$, and for different number of layers in $U$. For the QCNN, the max number of layers is $n_L=\lceil \log(n)\rceil$ (with $\lceil \cdot\rceil$ denoting the ceiling of a real-valued number), whereas for HEA we take $n_L=1,\cdots n$. For a QCNN (Fig.~\ref{fig:su4_hea_qcnn_maxchi}(top)), we can see that the maximum bond dimension is always kept within reasonable levels (as further evidenced from the large scale numerics performed in Fig.~\eqref{fig:deep_su4_hea}). Here, we can clearly see the effect of the ceiling in the number of layers, as when $n$ surpasses a power of $2$, the bond dimension exhibits small jumps. Then, for the HEA (Fig.~\ref{fig:su4_hea_qcnn_maxchi}(bottom)), we can surprisingly see that the bond dimension is always $4$ irrespective of the number of qubits, and the number of layers. Both of these results show that our TN methods are well behaved and scalable for the circuit topologies considered.

\subsubsection{Comparison with Monte Carlo techniques}

\begin{figure}
    \centering
    \includegraphics[width=1\columnwidth]{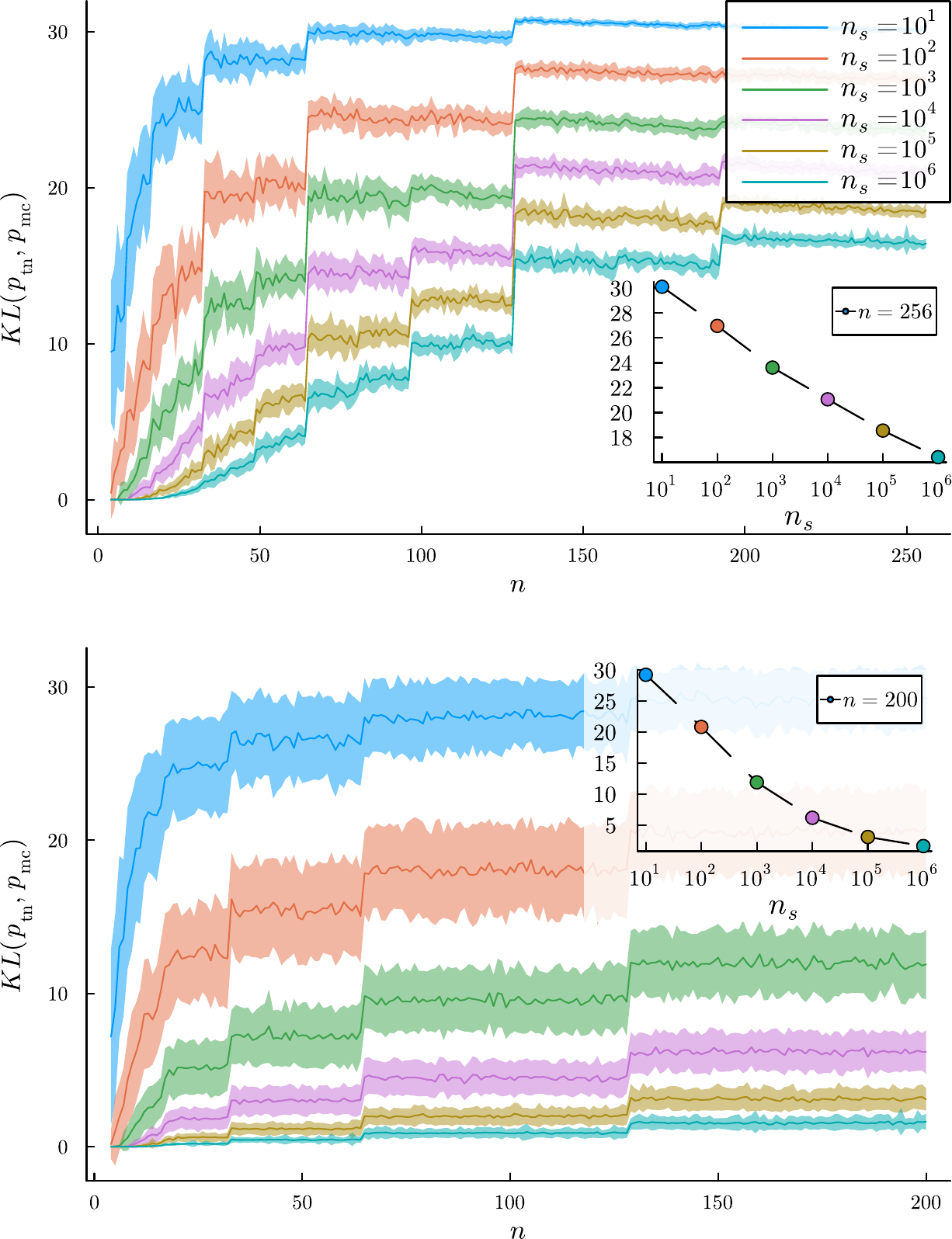}
    \caption{Scaling of Kullback-Leibler divergence between the purity distributions obtained via TNs (exact distribution) and via MC sampling for a QCNN (top) and an HEA (bottom) of different system sizes, and for different number $n_s$ of MC samples. The curves show the average (solid line) and standard deviations (ribbon) over 100 independent MC runs. The insets show the decaying of the KL divergence at a fixed system size (largest $n$ considered) as the number of MC samples increases.}
    \label{fig:mc_vs_tn_su4_loghea}
\end{figure}

As briefly discussed in the previous sections, a common approach in the literature to exactly estimate the $t=1,2$ moments is the use of  MC sampling for the Markov chain-like process~\cite{napp2022quantifying}. To illustrate the advantage of our TN method versus MC sampling, we have implemented the algorithm in~\cite{napp2022quantifying} to reconstruct the $k$-purity distribution via MC sampling for a QCNN and for an HEA, both with depth $n_L=\lceil \log(n)\rceil$. Denote as $p_{\rm tn}$ the ground-truth purity distribution obtained via our TN methods, and as $p_{\rm mc}$ the MC purity distribution. In Fig.~\ref{fig:mc_vs_tn_su4_loghea} we plot the scaling of the Kullback-Leibler (KL) divergence $KL(p_{\rm tn}, p_{\rm mc}) = \sum p_{\rm tn}\log (p_{\rm tn}/p_{\rm mc})$ between those two distributions, as the number of MC samples $n_s$ increases from $n_s=1$ to $n_s=10^6$. 
As we can see  in Fig.~\ref{fig:mc_vs_tn_su4_loghea}(top), when $U$ is a QCNN, MC fails to accurately reproduce the purity distributions even at $n_s=10^6$ samples. Moreover, we can see that as the number of qubits increases past a power of two, which implies an increase in the number of layers, the performance of MC decreases. In fact, as shown in the inset of this panel, at $n=250$ qubits, an exponential increase in shots, leads to what appears to be a sub-linear improvement in the KL divergence. This fact implies that a super-exponential number of MC samples might be needed to decrease the KL divergence by a constant amount. Then, as we can see in Fig.~\ref{fig:mc_vs_tn_su4_loghea}(bottom), for an HEA, while the KL divergence is smaller, indicating that MC can indeed lead to better purity distributions, the same phenomena occurs. Namely, exponentially increasing the number of measurements leads to smaller and smaller improvements in the KL divergence. 

Taken together, these results show that our TN can significantly outperform standard MC sampling techniques.

\subsubsection{Entanglement properties of the vectorized MPSs}

In this section we showcase a novel dimension of analysis enabled from our TN approach. First, let us recall the (obvious) fact that as $|O^{\otimes t}\rangle \rangle$ evolves through the gates of the $P$-net, it will always be an un-normalized vector in $\mathbb{R}^{\otimes n}$. As such, if we normalize it, we can consider that the MPS now represents a quantum state. Combining this realization with the fact that this MPS keeps a  small bond dimension as it propagates  through the layers of the $P$-net, allows for studying its entanglement and entropic properties.

Indeed, given a well-behaved MPS one can compute the entanglement properties of some reduced state~\cite{orus2014practical,biamonte2017tensor}. For instance, given a bipartition of the set of $n$ MPS indexes as $\{I,\Bar{I}\}$, we can compute  the entropy of entanglement, i.e., the Von Neumann entropy, given by 
\begin{equation}
    S(I) = - \Tr [\rho_{I} \log (\rho_{I})]\,,
\end{equation}
where $\rho_I$ is the reduced state of the MPS obtained by tracing out the qubits whose indexes are in $\Bar{I}$. Similarly, we can also compute the second R\'enyi entropy
\begin{equation}
    S_2(I) =- \frac{1}{2}\log (\Tr [\rho_{I}^2])\,.
\end{equation}
Notice that the computational cost of computing $S_2$ scales as $\chi^5$, implying that we need $\chi$ to be small for this calculation to be efficient. For instance, we can see from Fig.~\ref{fig:su4_hea_qcnn_maxchi} that we can readily compute $S_2$ for the HEA circuit, whereas the same does not apply to the QCNN.

In what follows we will define several sets of indexes leading to reduced states of interest. First, we define the simplest set $Q_j=\{j\}$, containing only index $j$, so that $\rho_{Q_j}$ is the reduced density matrix on the $j$-th qubit, and $S(Q_j)$ quantifies the entanglement between the $j$-th qubit and the rest. Next, we define the set of  ``edge'' qubits indexes $E_j=\{1,2,\cdots j\}$, so that $\rho_{E_j}$ is the reduced state on all the qubits from $1$ to $j$.  Finally, we define as $M_j = \{n/2 -j,n/2 -j+1,\cdots, n/2 +j]$ as the set of ``middle'' qubit indexes, spanning those that spread from the middle qubit  outwards with radius $j$.

\begin{figure}
\centering
\includegraphics[width=1\columnwidth]{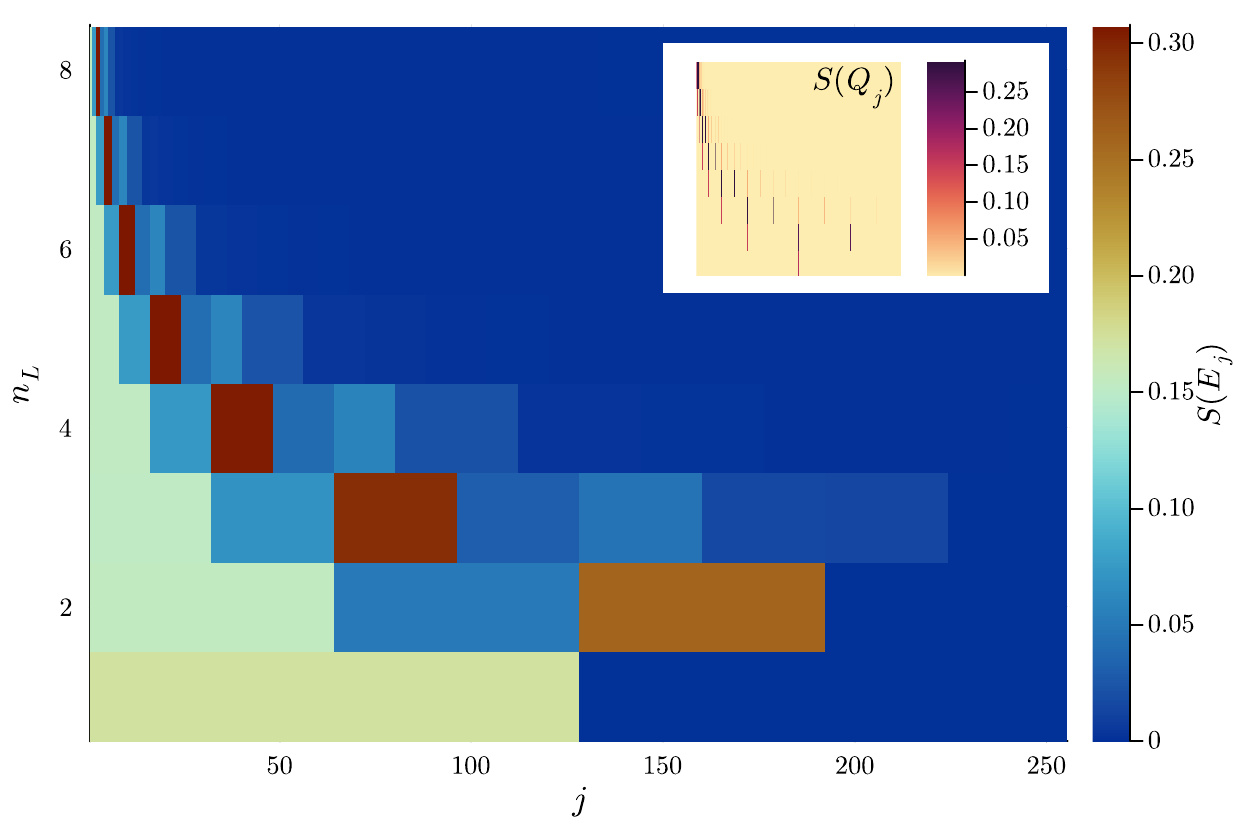}
\hfill
\centering
\includegraphics[width=1\columnwidth]{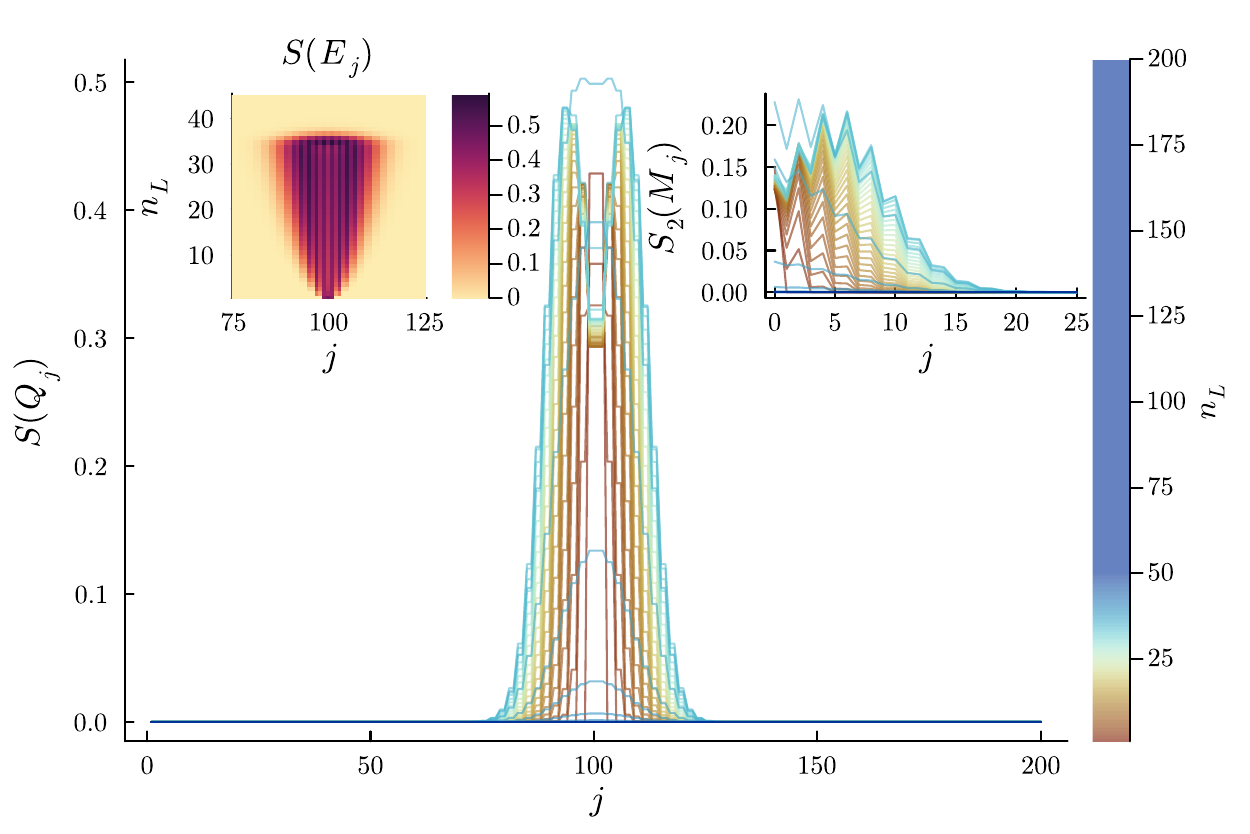}
\caption{Entropic measures of entanglement for the vectorized MPS representation of the vectorized measurements as it propagates through the $P$-net associated to a QCNN (top) and a HEA (bottom). Here, $S$ and $S_2$ are the Von Neumann and second R\'enyi entropies, respectively. The entropies of   $Q_j$ quantify the entanglement between qubit $j$ and the rest, while $E_j$ ($M_j$) the one between the edge (middle) qubits and the rest (see the main text). }
\label{fig:entanglement_su4}
\end{figure}

In Figure~\ref{fig:entanglement_su4} we present results for a numerical study of the entanglement properties of (the normalized) $\widehat{\tau}^{(2)}|O^{\otimes t}\rangle \rangle$ for the $P$-nets of a QCNN (top) and HEA (bottom). 
For the QCNN panel the system size is set to $n=256$, thus forcing the number of layers to be $n_L=8$, and the measurement operator is chosen to be $O=Z_1$. We show the heatmap of $S(E_j)$ for cuts at each possible index $j$ and increasing number of layers. The inset shows an analogous plot for $S(Q_j)$. 
In the HEA panel we use $n=200$ and a total number of layers $n_L=n=200$, while the measurement operator is $O=Z_{n/2}$. This time the main plot shows $S(Q_j)$, where we color-coded the increasing number of layers. The two insets at the left and right depict, respectively, the heatmap of $S(E_j)$ and a color-coded plot of $S_2(M_j)$.  Notice that the right inset shares the same color-coding of $n_L$ with the main plot.

As we can see from both panels in Figure~\ref{fig:entanglement_su4} , all the correlations are initially stored in the causal cone of $O$, to then quickly get washed out and become localized to the first few sites (QCNN) or completely vanish (HEA). In the QCNN case, we can particularly appreciate the switching on of correlations in each of the layers' support and their later decay. For the HEA, the fact that the correlations disappear with depth can be easily interpreted by recalling that, as discussed in the previous sections, a deep HEA forms a $2$-design over $U(2^n)$. One can then use Weingarten calculus to find the MPS to which $\widehat{\tau}^{(2)}|O^{\otimes t}\rangle \rangle$ converges to, and prove that it corresponds to a quantum state whose distance from a product state is exponentially small in $n$. Namely, as shown in the appendix, we can find that for a deep HEA which forms a $2$-design over $U(2^n)$, one has that $\forall j$ 
\begin{equation}
    \rho_{Q_j}= \frac{1}{4^n - 1} \begin{pmatrix}
        4^{n-1} - 1 & \sqrt{3}(4^{n-1} - 1) \\
        \sqrt{3}(4^{n-1} - 1) & 3\,4^{n-1}\\
    \end{pmatrix}\,,
\end{equation}
whose associated eigenvalues are
\begin{equation}
    \left\{\frac{1}{2} \left(\frac{\sqrt{-5\ 4^n+16^n+13}}{4^n-1}+1\right), \frac{1}{2}-\frac{\sqrt{-5\ 4^n+16^n+13}}{2 \left(4^n-1\right)}\right\}\nonumber
\end{equation}
which leads to $S(Q_j)=0$, as the eigenvalues converge to $\{1,0\}$, in the $n\rightarrow \infty$ limit.

\subsection{Local gates sampled from $G = O(4)$}\label{sec:O-results}

In this section we will use our TN machinery to study the second moments of the expectation value for the case when $U$ is a  quantum circuits composed of two-qubit random gates ($d=2$ and $k=2$) drawn i.i.d. from the fundamental representation of $O(4)$. Following the same line of thought as that used to determine the $P$-gates for the $U(4)$ examples  above, we need to find a local basis for the $P$-net. As indicated by Theorem~\ref{theo:fundamental}, the local dimension of each $P$-gate legs is $3$. In particular, we can use the following orthogonal local basis  $\{|\id\rangle\rangle,|{\rm S}\rangle\rangle,  |{\rm B}\rangle\rangle\}$, where $|{\rm \id}\rangle\rangle, |{\rm S}\rangle\rangle$ are the same operators appearing in the $U(4)$ basis and where $|{\rm B}\rangle\rangle = |X\otimes X - Y\otimes Y + Z\otimes Z\rangle\rangle$. This choice leads to $9\times 9$ ($3\times 3\times 3\times 3$ tensor) $P$-gates reading
\fontsize{7.5}{8}\selectfont
\begin{equation}\label{eq:matrix-P-so-0}
    P=\begin{pmatrix}
1 & 0 & 0 & 0 & 0 & 0 & 0 & 0 & 0 \\
0 & {7/36} & {1/36} & {7/36} & {11/18} & {1/6} & {1/36} & {1/6} & -{1/18} \\
0 & {1/36} & {7/36} & {1/36} & -{1/18} & {1/6} & {7/36} & {1/6} & {11/18} \\
0 & {7/36} & {1/36} & {7/36} & {11/18} & {1/6} & {1/36} & {1/6} & -{1/18} \\
0 & {7/36} & {1/36} & {7/36} & {11/18} & {1/6} & {1/36} & {1/6} & -{1/18} \\
0 & 0 & 0 & 0 & 0 & 0 & 0 & 0 & 0 \\
0 & {1/36} & {7/36} & {1/36} & -{1/18} & {1/6} & {7/36} & {1/6} & {11/18} \\
0 & 0 & 0 & 0 & 0 & 0 & 0 & 0 & 0 \\
0 & {1/36} & {7/36} & {1/36} & -{1/18} & {1/6} & {7/6} & {1/6} & {11/18}
\end{pmatrix}\,.
\end{equation}
\normalsize
More details on the derivation of the $O(4)$ $P$-gates can be found in the appendices.

Importantly, we can see from Eq.~\eqref{eq:matrix-P-so-0} that the $P$-matrix contains both positive and negative signs. This will lead to a Markov chain-like process matrix with non-negligible negative entries, indicating that a Monte Carlo simulation is not available. Thus, all the results presented in this section, which are obtained Via our TN methods, cannot be readily reproduced via Monte Carlo sampling techniques. 

We note that instead of presenting plots for the $k$-purities in random quantum circuits with orthogonal gates (such as those presented in the previous section), we instead opt  for the more interesting route of  using our TN methods to study the anticoncentration phenomenon on these circuits.

\subsubsection{Anticoncentration of shallow orthogonal local random quantum circuits}

Anticoncentration \cite{dalzell2022randomquantum,hangleiter2018anticoncentration} captures how much the probabilities of obtaining different outcomes when measuring a quantum circuit are similar to each other, i.e, how much the outcome probability distribution of the quantum circuit resembles that of a uniform distribution. Particularly, a random circuit architecture is said to be anticoncentrated if the probability of any measurement outcome is, at most, a constant factor larger than the uniform value. Quantitatively, anticoncentration can be studied by computing the so-called collision probability averaged over the randomly chosen circuits~\cite{dalzell2022randomquantum}
\begin{equation}
    Z = \mathbb{E}_U \left[ \sum_{x\in [d]^n} p_U(x)^2 \right] \, ,
\end{equation}
where $x$ is a computational basis state of the $n$-qudits circuit $U$, and $p_U(x)$ is the probability of measuring it as outcome. If the circuit anticoncentrates, each of the sum terms has to be at most a constant factor greater than the uniform probability $d^{-n}$. Hence one obtains the anticoncentration bound as the condition of the existence of a factor $\alpha$ independent of $n$ such that $Z<\alpha d^{-n}$. For instance, it has been analytically shown that the outcome probabilities of one-dimensional HEAs, such as those depicted in Fig.~\ref{fig:topologies}(b), with local random $U(4)$ gates anticoncentrate at $n_L=\Theta(\log(n))$ depth~\cite{dalzell2022randomquantum}. We here analyze if a  similar phenomenon will occur if the HEA gates are sampled from $O(4)$ rather than $U(4)$.

First, let us note that given a circuit $U$ with local Haar random $O(4)$ gates, all the computational basis states will be equiprobable (as long as at least one gate acts on each qubit). This follows from the fact that the bit-flip transformation $X$ is in $SO(4)$. 
Hence, we can study the anticoncentration by analyzing the simplified quantity $Z=d^n \mathbb{E}_U[p_U(\ket{0}^n)^2]$. Moreover, we can readily obtain the value $Z_H$ at which $Z$ will converge in the limit of infinite depth, i.e., when $U$ becomes a random sample from the global $O(2^n)$, which reads $Z_H = 3/(2^n+2)$~\cite{garcia2023deep}. 

\begin{figure}
    \centering
    \includegraphics[width=1\columnwidth]{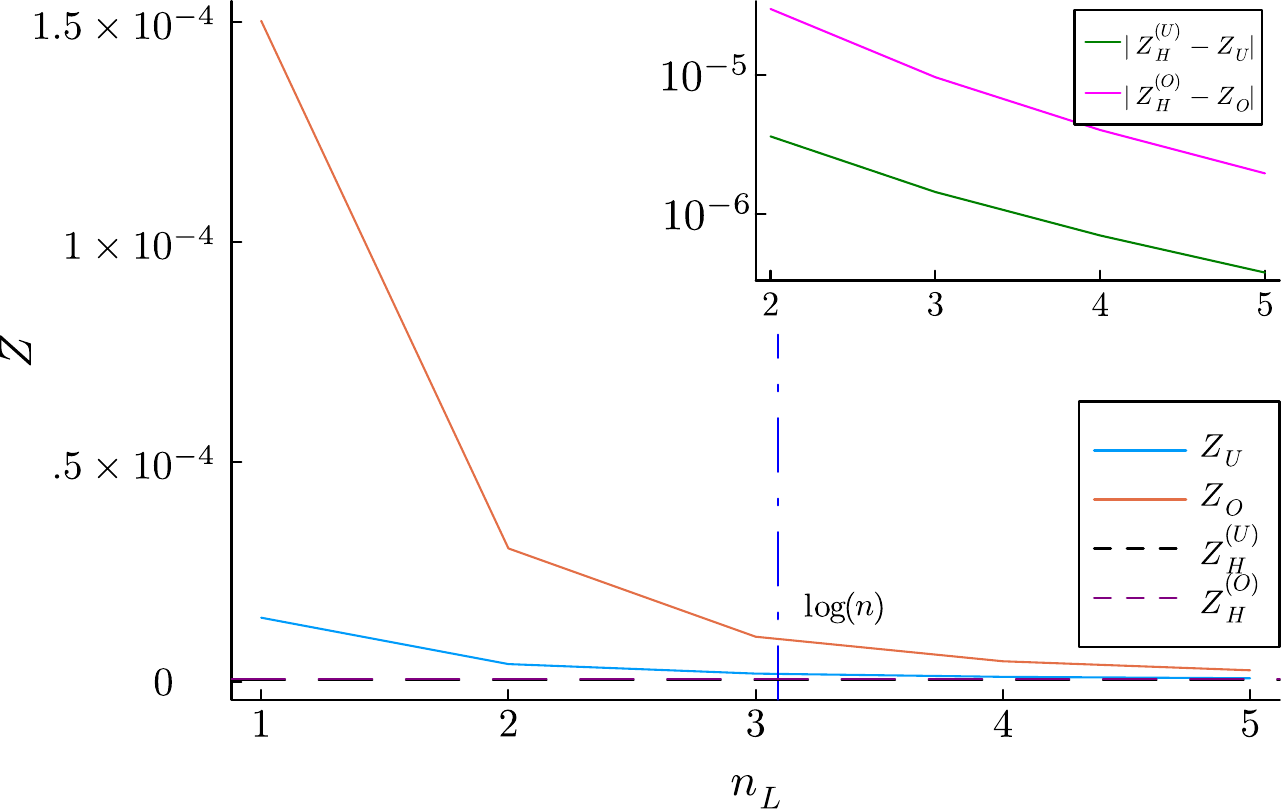}
    \caption{Averaged collision probability $Z_{U/O}$ for a one-dimensional HEA composed of local random gates sampled i.i.d. from  $O(4)$ (orange line) and $U(4)$ (blue line) and acting on $n=22$ qubits, as a function of the number of layers $n_L$. The dashed lines shows $Z_H^{(U/O)}$, the asymptotic value of $Z_{U/O}$ as $n_L$ tends to infinity. The inset shows the behavior in a neighborhood of $n_L=\log(n)$.}
    \label{fig:so4_anticoncentration}
\end{figure}

In Figure~\ref{fig:so4_anticoncentration} we show the behavior of $Z$ as a function of the number of layers $n_L$ for a circuit acting on $n=22$ qubits. For comparison, we also show the same behavior for the HEA with local gates samples from $U(4)$ (i.e., the case analyzed in \cite{dalzell2022randomquantum}). Here, we can see that as the number of layers increases, the average collision probability quickly converge to their infinite depth limit. We can clearly see that for all values of $n_L$ the value of $Z$ is smaller for a circuit with unitary random gates (which is expected due to its larger expressive power).  Nonetheless, it seems like random quantum circuits with orthogonal gates still anticoncentrate at  logarithmic depth.  

\section{Discussion}

Tensor networks have emerged as a crucial tool in the realm of statistical physics and condensed matter physics. The significance of tensor networks lies in their ability to represent complex correlations efficiently, allowing for the exploration of evolutions in vector spaces that are otherwise computationally intractable. Importantly, unlike Monte Carlo techniques, which struggle with process matrices exhibiting a negative sign, tensor networks offer an alternative approach that sidesteps this issue. 

In this work, we follow suit in this line of thought and ask the question: ``\textit{Can we use tensor networks, instead of Monte Carlo sampling, to exactly compute the moments of expectation values obtained from  quantum circuits composed of local random gates?}'' Our question is motivated from the fact that the moments can be expressed --via vectorization-- as the inner product between two vectors and a Markov chain-like process matrix. Indeed, we not only show that this approach is mathematically sound, but also advantageous over standard Monte Carlo sampling techniques. 

Our first main contribution is a description of the mathematical framework needed to exactly compute quantities such as $\mathbb{E}_U[\Tr[U\rho U\ad O]^t]$ via tensor networks. The formalism is presented in a general way, allowing for local gates acting on different number of qubits, and being uniformly sampled from different local groups. We then use representation theoretical tools to derive theoretical results which analyze the local dimension of the tensor, as well as present bounds for the maximum bond dimension of the matrix product states that deep circuits can produce. Next, we showcase our method for estimating the second moment of two-types of quantum neural networks with local two-qubit gates sampled from the fundamental representations of $U(4)$ and $O(4)$. Here, we illustrate that our methods can efficiently tackle circuits acting on thousands of qubits, and composed of thousands of gates. These results also illustrate the fact that tensor networks can significantly outperform Monte Carlo simulations in terms of the desired estimation accuracy, but also by being able to tackle tasks where Monte Carlo would exhibit sign problems (local orthogonal circuit anticoncentration), or not be appropriate (computing entropic properties of the MPS).

While our numerical simulations have demonstrated the capabilities of the use of TN to study random quantum circuits, there are also many open questions and new research directions opened by our results. For instance, we naturally expect that our proposed techniques will encounter issues for problems with complicate topologies, or when the local tensor dimensions leads to prohibitively large bond dimensions. In this context, we note that it is not clear to us how to derive general bounds for the local tensor dimensions, and we expect that a rigorous mathematical analysis will be needed on a case-by-case basis (see for instance our results in Theorem~\ref{theo:-free-fermion} for free-fermion circuits with gates sampled from $SO(4)\simeq SU(2)\oplus SU(2)$). We leave for future work a detail exploration of this optimal local basis questions, and we hope that representation theoretic tools can be used to make progress in this regard. Then, we also note that the representation of the circuit moment operator itself could be used to learn properties of the quantum circuit independently of the initial state and measurement operator. For instance, we could use density matrix renormalization group techniques to obtain its eigenvalues, and thus be able to predict the number of layers needed for the circuit to become a $t$-design over $G_U$. On a similar note, it is worth highlighting that having access to a matrix product state representation of $\widehat{\tau}^{(t)}|\rho^{\otimes t}\rangle \rangle$   allows a whole new dimension of random quantum circuit analysis, such as the study of the entanglement and entropic properties of this quantum state. Indeed, we can already see that while the entanglement present in the MPS for unitary circuits remains small (as evidenced from the small bond dimension), such is not the case for orthogonal circuits. In the latter case, the $P$ matrices contain negative signs, and these appear to lead to more entangled MPSs~\cite{chen2024sign}. While we have not explored in this work the connection between the MPS entanglement, the dynamics in the random circuit, and their relation to the representation theoretical properties of the TN, we envision that these will be fruitful directions of research. Finally, we note that the proposed tensor network formalism can be readily applied to random quantum circuits with intermediate measurements, thus enabling the study of monitored random dynamics  and measurement-induced criticality~\cite{jian2020measurement,bao2020theory,nahum2021measurement,fisher2023random}. As such, given the versatility of our proposed techniques,  we envision that tensor networks  will quickly become a standard  tool in the toolbox of quantum information scientist studying and working with circuits composed of random local gates. 

\textit{Note added:} A few days before our manuscript was uploaded as a preprint, we became aware of the work~\cite{hu2024demonstration}, which presents a method for treating the $P$-net as a TN similar to one presented here. We note that while some of the techniques in Ref.~\cite{hu2024demonstration} are similar to ours, that work only considered the case of local unitary gates. As such, the extension to local orthogonal or free-fermionic gates introduced here is, to our knowledge, completely novel.

\medskip

\section{Acknowledgments}

We sincerely thank Hsin-Yuan (Robert) Huang, Martin Larocca, Diego Garc\'ia-Mart\'in, and Francesco Caravelli for useful discussions. The authors were  supported by the Laboratory Directed Research and Development (LDRD) program of Los Alamos National Laboratory (LANL) under project numbers 20230049DR (P.Braccia and L.C),  and 20230527ECR (P.Bermejo and M.C.). P. Bermejo acknowledges DIPC for constant support. M.C. was also initially supported by LANL ASC Beyond Moore’s Law project.

\bibliography{quantum}

\newpage
\appendix

\section{Proof of Theorem 1 and Proposition 1}

In what follows, we present a joint proof for Theorem~\ref{theo:fundamental} and Proposition~\ref{prop:bond-dimension}. For simplicity we start by considering the case when all the local gates are sampled from  $G=U(d^k)$, and then study the case of $G=O(d^k),SP(d^k/n)$.  

\begin{proof}
Let $V$ be a unitary acting on $k$-qudits.   Then, we know from the Schur-Weyl duality that the $t$-th order commutant of the fundamental representation of the unitary group $U(d^k)$ is spanned by the subsystem permuting representation of the symmetric group $S_t$. Specifically, the subsystem permuting representation of a permutation $\sigma\in S_t$ is 
\begin{equation}\label{eq:rep-S_k-main}
P(\sigma)=\sum_{i_1,\dots,i_k=1}^{d^k} |i_{\sigma^{-1}(1)},\dots,i_{\sigma^{-1}(k)} \rangle\langle i_1,\dots,i_k|\,,
\end{equation} 
from which it can be verified that $[V^{\otimes t},P(\sigma)]=0$ for any $V\in U(d^k)$ and for all $\sigma\in S_t$. 

Next, let us note that any permutation of the $t$-copies of the $k$-qudit Hilbert space can be expressed as  
\begin{equation}
    P(\sigma)=\bigotimes_{\mu=1}^k P_\mu(\sigma)\,,
\end{equation}
where $P_\mu(\sigma)$ permutes the copies of the $\mu$-th qudits. From this, it follows that  in the vectorized formalism we can express
\begin{equation}
    |P(\sigma)\rangle \rangle=\bigotimes_{\mu=1}^k |P_\mu(\sigma)\rangle\rangle\,,
\end{equation}
indicating that $|P(\sigma)\rangle \rangle$ is an MPS of bond dimension one. Thus, since there are $t!$ operators in $S_t$, and since each one of them corresponds to a bond dimension one MPS, we can always construct the $P$ matrix for the moment operator in a tensor product basis where each $P(\sigma)$ is a basis element, leading to $P$  being  a square matrix of dimension up to $(t!)^2\times (t!)^2 $. 

From here, it is not hard to see that if the group is $G_U=U(d^n)$, then since  $\widehat{\tau}^{(t)}$ will be a projector onto ${\rm comm}^{(t)}(U(d^n))$, then for any $A\in \BC(\HC^{\otimes t})$, the vectorized operator $\widehat{\tau}^{(t)}|A\rangle\rangle$ will be a linear combination of the $ |P(\sigma)\rangle \rangle$. Since there are $t!$ of them, and since they are all bond dimension-one MPSs, the maximum bond dimension of the vector $\widehat{\tau}^{(t)}|A\rangle\rangle$ is $t!$.

Next, let us consider the case of $O(d^k)$ and $Sp(d^k/2)$ (assuming $d$ is even). Here, we know from the Schur-Weyl duality that the $t$-th order commutant of the fundamental representation of the orthogonal group $O(d^k)$ is spanned by the Brauer algebra $B_{2t}$~\cite{larocca2022group}.  Given that the $t$-th order commutant of $Sp(d^k/2)$ follows the same structure as the Brauer algebra, we will focus on the orthogonal group, with the proof of the symplectic group following similarly. 

Here, we only need to know that for any element $\sigma$ of the Brauer algebra it follows that 
\begin{equation}
    |P(\sigma)\rangle \rangle=\bigotimes_{\mu=1}^k |P_\mu(\sigma)\rangle\rangle\,,
\end{equation}
meaning that the $P$ matrices can also always be constructed in a basis where each $|P(\sigma)\rangle\rangle $ is a basis element. Hence, since there are $\frac{(2t!)}{2^tt!}$ such elements, the $P$  matrix will be  square and of dimension up to $\left(\frac{(2t!)}{2^tt!}\right)^2\times \left(\frac{(2t!)}{2^tt!}\right)^2 $. With a similar reasoning, the maximum bond dimension of any operator projected by the $t$-th moment operator of a deep circuit with $G_U=O(d^n)$ will be $\frac{(2t!)}{2^tt!}$.

\end{proof}

\section{Proof of Theorem 2 and Proposition 2}

In this section we present a proof for Theorem~\ref{theo:-free-fermion} and Proposition~\ref{prop:bond-dimension-ideals} an $n$-qubit circuit composed of free-fermionic gates~\cite{diaz2023showcasing}. First, let us recall a few basic properties of the free-fermionic representation of the group $SO(2n)$. To begin, we recall that the  Lie algebra $\mathfrak{so}(2n)$ associated to this group has a basis given  by~\cite{zimboras2014dynamic,kokcu2022fixed,wiersema2023classification}
 \begin{equation} \label{sup-eq:dla}
    \mathfrak{so}(2n)={\rm span}_{\mathbb R}i\{Z_i,\widehat{X_iX_j},\widehat{Y_iY_j},\widehat{X_iY_j},\widehat{Y_iX_j}\}_{1\leq i<j\leq n} \,,
 \end{equation}
 \normalsize
where we use the notation $\widehat{A_iB_j}=A_iZ_iZ_{i+1}\cdots Z_{j-1}B_j$. To work with this algebra, it is convenient to define the $2n$ Majorana operators~\cite{jozsa2008matchgates}
\begin{equation}\begin{split} 
    c_1&=X\id\dots \id,\; c_3= ZX\id\dots \id, \;\dots,\; c_{2n-1}=Z\dots Z X \\
        c_2&=Y\id\dots \id,\; c_4= ZY\id\dots \id, \;\dots, \;\; c_{2n}\;\;\;=Z\dots Z Y\,, \end{split}\nonumber
\end{equation}
which are proportional to Pauli operators that satisfy the anti-commutation relation 
\begin{equation}\label{sup-eq:majo-anit-comm}
    \{c_\mu,c_\nu\}=2\delta_{\mu\nu}\;\;\;\; \mu,\nu=1,\dots ,2n\,.
\end{equation}
In this basis, all the elements of $\mathfrak{so}(2n)$ in~\eqref{sup-eq:dla} can be expressed as the product of two Majoranas. That is,
\begin{equation}
    \mathfrak{so}(2n) = {\rm span}_{\mathbb{R}}\{c_\mu c_\nu\}_{1\leq\mu<\nu\leq2n}\,.
\end{equation}

Next, we note that the adjoint action of $SO(2n)$ over the operator space $\BC(\HC)$ is in itself a representation of the Lie group, and hence induced a decomposition of $\BC(\HC)$ into invariant subspaces, or group-modules. In particular, one can prove that $\BC(\HC)$ can be decomposed into modules as~\cite{diaz2023showcasing}
\begin{equation} 
\BC(\HC)=\bigoplus_{\kappa=0}^{2n} \mathcal{B}_{\kappa}\,,
\end{equation}
where each  $\mathcal{B}_\kappa$ is the linear space, of dimension $\binom{2n}{\kappa}$, spanned by a basis of products of $\kappa$ distinct Majoranas. As such, we know that given any $A\in\BC_\kappa$, then $VAV\ad \in\BC_\kappa$ for any $V\in SO(2n)$. 

Having characterized the Lie group and its associated Lie algebra, let us proceed to study their $t$-th fold commutants. First, for $t=1$, we know that the only symmetry of the group is the fermionic parity operator $X^{\otimes n}=(-i)^{n}c_1c_2\cdots c_{2n}$, so that
\begin{equation}
    {\rm comm}^{(1)}(SO(2n))={\rm span}_{\mathbb{C}}\{\id,Z^{\otimes n}\}\,.
\end{equation}
Then, an orthonormal basis of size $2(2n+1)$ for ${\rm comm}^{(2)}(SO(2n))$ is given by ~\cite{oszmaniec2022fermion,diaz2023showcasing}
\begin{equation}
\label{sup-eq:comm-basis} 
    \begin{split}
        &Q_\kappa^0 = \NC_\kappa \sum_{\vec{s}\in\binom{[2n]}{\kappa}} c^{\vec{s}}\otimes c^{\vec{s}} \,,\\ &Q_\kappa^1 = \NC_\kappa\,  (-i)^n i^{\kappa\; {\rm mod}\;2} \sum_{\vec{s}\in\binom{[2n]}{\kappa}} (-1)^{\pi(\vec{s})}c^{\vec{s}}\otimes c^{\vec{\bar{s}}} \,,
    \end{split}
\end{equation}
for integers $0\le \kappa\le 2n$, and $\NC_\kappa =  \left(2^n \sqrt{\binom{2n}{\kappa}}\right)^{-1}$. Here, $c^{\vec{s}}$ denotes a product of $\binom{[2n]}{\kappa}$ distinct Majoranas, as labeled by the index $\vec{s}$. 

With these results in mind, let us first prove Theorem~\ref{theo:-free-fermion}.

\begin{proof}
    Let $V$ be a two-qubit gate sampled from the Haar measure of the free-fermionic representation of the group $SO(4)$. Then, according to Eq.~\eqref{sup-eq:comm-basis},  a basis for ${\rm comm}^{(2)}(SO(2n))$ contains 10 elements. For simplicity, let us first consider the first five parity respecting elements $Q_\kappa^0$ which we can express in the Pauli basis as
    \small
    \begin{align}
        Q_0^0=&\id\id\otimes \id\id\nonumber\\
        Q_1^0=&X\id\otimes X\id+Y\id\otimes Y\id+ZX\times ZX+ZY\otimes ZY\nonumber\\
        Q_2^0=&Z\id\otimes Z\id+\id Z\otimes \id Z+XX\otimes XX+XY\otimes XY\nonumber\\
        &+YY\otimes YY+YX\otimes YX\nonumber\\
        Q_3^0=&\id X\otimes \id X+\id Y \otimes \id Y+XZ\otimes XZ+YZ\otimes YZ\nonumber\\
        Q_4^0=&ZZ\otimes ZZ\nonumber
    \end{align}
\normalsize
Here, a Pauli operator $AB\otimes CD$ is read as: $A$ and $B$ act on the first and second qubits, respectively,  of the first copy of the Hilbert space; while $C$ and $D$ act on the first and second qubits, respectively,  of the second copy of the Hilbert space. By vectorizing this operator, and re-ordering the indexes of the tensor product to $AB\otimes CD\rightarrow |AC\rangle\rangle\otimes|BD\rangle\rangle $, and omitting redundancies as $|A\rangle\rangle|B\rangle\rangle\equiv |AA\rangle\rangle\otimes|BB\rangle\rangle$ we find
    \begin{align}
        |Q_0^0\rangle\rangle=&|\id\rangle\rangle |\id\rangle\rangle\nonumber\\
        |Q_1^0\rangle\rangle=&|X+Y\rangle\rangle|\id\rangle\rangle+|Z\rangle\rangle|X+Y\rangle\rangle\nonumber\\
        |Q_2^0\rangle\rangle=&|Z\rangle\rangle |\id\rangle\rangle
        +|\id\rangle\rangle |Z\rangle\rangle+|X+Y\rangle\rangle|X+Y\rangle\rangle\nonumber\\
        |Q_3^0\rangle\rangle=& |\id\rangle\rangle|X+Y\rangle\rangle+ |X+Y\rangle\rangle|Z\rangle\rangle\nonumber\\
       | Q_4^0\rangle\rangle=&|Z\rangle\rangle|Z\rangle\rangle\nonumber
    \end{align}
which shows that the local commutant's basis for the vectorized  (parity respecting) $2$-nd order commutant of $SO(4n)$ is composed of three elements $\{|\id\rangle\rangle,|Z\rangle\rangle,|X+Y\rangle\rangle\}$. A similar procedure can be performed on the rest of the elements $Q_\kappa^1$, which will lead to three additional basis vectors. Therefore,  the $P$ matrix for $\widehat{\tau}_{SO(4)}^{(2)}$ will be of dimension $9\times 9$ if either $\rho$ or $O$ have fixed fermionic parity (as we only need to keep the basis elements coming from the $Q_\kappa^0$ operators), while it will be of size $36\times 36$ for generic states and observables.
\end{proof}

Next, let us provide a proof for Proposition~\ref{prop:bond-dimension-ideals}.

\begin{proof}
First, let us recall that we assume that the random circuit $U$  forms a $2$-design over $SO(2n)$. Then, we know that the circuit moment operator $\widehat{\tau}^{(2)}$ will be a projector onto the commutant ${\rm comm}^{(2)}(SO(2n))$ spanned by the $2(2n+1)$ elements in Eq.~\eqref{sup-eq:comm-basis}. If $A$ is a Pauli operator belonging to $i\mathfrak{so}(2n)$, then we know that $A^{\otimes 2}$ can only have non-zero overlap with the Casimir operator $Q^0_2$. A direct calculation reveals that $\Tr[Q^j_\kappa A^{\otimes 2}]=\delta_{j,0}\delta_{\kappa,2}\langle\langle Q^0_2|A\rangle\rangle= \delta_{j,0}\delta_{\kappa,2} 2^n \NC_\kappa $. More precisely, we have that $2^n \NC_2=1/\sqrt{\binom{2n}{\kappa}}\in\Omega(1/\poly(n))$.

From the previous, find
\begin{align}
    \widehat{\tau}^{(2)}|A^{\otimes2}\rangle \rangle&=\sum_{j=0,1}\sum_{\kappa=0}^{2n } |Q^j_\kappa\rangle \rangle \langle\langle Q^j_\kappa|A^{\otimes t}\rangle \rangle\nonumber\\
    &=\frac{1}{\sqrt{\binom{2n}{\kappa}}}|Q^0_2\rangle \rangle\,.
\end{align}
Hence, the maximum bond dimension of $\widehat{\tau}^{(2)}|A^{\otimes2}\rangle \rangle$ will be equal to that of $|Q^0_2\rangle \rangle$. Denoting as $S=\{Z_i,\widehat{X_iX_j},\widehat{Y_iY_j},\widehat{X_iY_j},\widehat{Y_iX_j}\}$ the basis for $i\mathfrak{so}(2n)$, we know that we can express $Q^0_2$ as $\sum_{ P\in S}P\otimes P$. A direct vectorization of this operator, along with a basis reordering and elimination of redundancies leads to  
\begin{equation}
    |Q^0_2\rangle \rangle\propto \sum_{j=1}^{n}|Z_j\rangle\rangle+\sum_{1\leq i<j\leq n}|\widehat{(X_i+Y_i)(X_j+Y_j)}\rangle\rangle\,,
\end{equation}
where 
\begin{equation}
    |\widehat{(X_i+Y_i)(X_j+Y_j)}\rangle\rangle=|(X_i+Y_i)Z_{i+1}\cdots Z_{j-1}(X_j+Y_j)\rangle\rangle\,.\nonumber 
\end{equation}
From here, we need to show that this an MPS of bond dimension $\chi=3$. For this purpose, we use  the following lemma.

\begin{lemma}\label{lem:MPS}
    The state
$$
\ket{\psi} = \sum_{i = 1}^n \ket{0\ldots01_i0\ldots0} + \sum_{1 \leq i < j \leq n} \ket{0\ldots02_i1\ldots12_j0\ldots0}
$$
can be described by the following (bond dimension $\chi=3$) MPS:
\begin{equation}
\ingr{1}{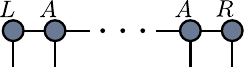} \ ,
\end{equation}
where
\begin{equation}
\begin{split}
\ingr{1}{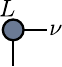} \  & = \
\left[ \begin{matrix}
\ket{0} & \ket{1} & \ket{2}
\end{matrix} \right]_{\nu} \ , \\
\ingr{1}{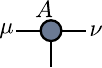} \ & = \ 
\left[\begin{matrix}
\ket{0} & \ket{1} & \ket{2} \\
0 & \ket{0} & 0 \\
0 & \ket{2} & \ket{1}
\end{matrix} \right]_{\mu,\nu} , \
\ingr{1}{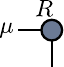} \ = \ \left[
\begin{matrix}
\ket{1} \\
\ket{0} \\
\ket{2}
\end{matrix} \right]_{\mu}
\end{split}
\end{equation}
\end{lemma}
The proof of this lemma is direct as we have explicitly constructed the MPS.

Using Lemma~\ref{lem:MPS}, we conclude our proof as $|Q^0_2\rangle \rangle$ can be mapped to the state $\ket{\psi}$ in the lemma via the relabeling $\id\rightarrow 0$, $Z\rightarrow 1$ and $(X+Y)\rightarrow 2$, and therefore is an MPS of bond dimension $\chi=3$.

\end{proof}

\section{Efficient computation of the $k$-purities}

Computing the $k$-purities $p^{(k)}_{O}$ of Eq.~\eqref{eq:k-puritites} requires computing sums over exponentially large spaces, as the number of $n$-qubits Paulis $P_j$ with bodyness $|P_j|=k$ is $N_k = 3^k \binom{n}{k}$. Hence, obtaining these quantities by individually evaluating the Hilbert-Schmidt products with all the $4^N$ Paulis would lead to an exponential wall-time, making the method unpractical. However, we can directly obtain $p^{(k)}_{O}$ from the MPS representation of $\widehat{\tau}^{(2)}|O^{\otimes 2}\rangle \rangle$ by computing its overlap with an MPS $|\phi_k \rangle \rangle$ that effectively projects onto the subspace of all Paulis with given bodyness $k$. Since the maximum bond dimension of $|\phi_k \rangle \rangle$, turns out to be $\lfloor n/2 \rfloor +1$, with $\lfloor x \rfloor$ being the integer part of $x$, this computation is efficient.

In order to construct these projectors one first needs to understand how to extract the overlaps with the trivial $\id_j^{\otimes 2}$, and non-trivial $X_j^{\otimes 2},Y_j^{\otimes 2},Z_j^{\otimes 2}$ Paulis from the local ${\rm comm}^{(2)}(G_j)$ bases on each of the physical legs $j\in [n]$ of the MPS $\widehat{\tau}^{(2)}|O^{\otimes 2}\rangle \rangle$ at the end of the $P$-net. 
To this end, we will introduce two vectors $\vert s_j^I \rangle\rangle, \vert s_j^P\rangle\rangle$ defined by the following properties
\begin{equation}\label{eq:s_vecs_properties}
\begin{split}
    \langle\langle s_j^I \vert O^{\otimes 2} \rangle\rangle &= \frac{1}{4} \Tr_j \left[\id_j^{\otimes 2} O^{\otimes 2} \right]\,,\\
    \langle\langle s_j^P \vert O^{\otimes 2} \rangle\rangle &= \frac{1}{4} \Tr_j \left[(X_j^{\otimes 2} + Y_j^{\otimes 2} + Z_j^{\otimes 2}) O^{\otimes 2} \right]\,,
\end{split}
\end{equation}
where $\Tr_j$ denotes the partial trace over the $j$-th qubit. 
For example, in the cases studied in Secs.~\ref{sec:U-results}, where all the gates are drawn i.i.d. from $G = U(4)$, one finds, for the local basis $\{ |\id \rangle\rangle, |{\rm S}\rangle\rangle\}$ used therein, that the vectors $\vert s_j^I \rangle\rangle, \vert s_j^P\rangle\rangle$ exactly correspond to the basis elements $\vert s_j^I \rangle\rangle = \vert \id \rangle\rangle \,\, \forall j$ and $\vert s_j^P \rangle\rangle = \vert {\rm S} \rangle\rangle \,\, \forall j$. 

Once these vectors have been figured out, computing $p_O^{(k)}$ turns into a matter of computing the contractions of the MPS at the end of the $P$-net with all the possible $\binom{n}{k}$ ways we can distribute $k$ $\vert s_j^P \rangle\rangle$ and $n-k$ $\vert s_j^I \rangle\rangle$ on its physical legs. That is, the projectors $|\phi_k \rangle \rangle$ that we are looking for can be written in the following form
\begin{equation}\label{eq:phi_k_definition}
    \vert \phi_k \rangle\rangle = \sum_{c \in \Delta_k^n}\left(\bigotimes_{j\in c} \vert s_j^P\rangle\rangle \right) \otimes \left(\bigotimes_{\Bar{j} \notin c} \vert s_{\Bar{j}}^P\rangle\rangle \right)  \,,
\end{equation}
where we defined $\Delta_k^n$ to be the set containing all the $\binom{n}{k}$ combinations of $k$ indices from $[n]$. For example $\Delta_1^3 = \{(1), (2), (3)\}$, and $\Delta_2^3 = \{(1, 2), (1, 3), (2, 3)\}$. Notice that together, Eqs.~\eqref{eq:s_vecs_properties} and \eqref{eq:phi_k_definition} correctly lead to
\begin{equation}
    p_O^{(k)} = \langle\langle \phi_k \vert \widehat{\tau}^{(2)} \vert O^{\otimes 2}\rangle \rangle \,,\\
\end{equation}
as can be easily checked from the definition of $k$-purities in Eq.~\eqref{eq:k-puritites}.

Then, the following Lemma shows how to build $\vert \phi_k \rangle \rangle$ as an MPS of maximum bond dimension $\chi=\lfloor n/2 \rfloor +1$.
\begin{lemma}\label{lem:phi_k}
    the vector $\vert \phi_k \rangle \rangle$ can be written as the following MPS:
\begin{equation}
\ingrp{1}{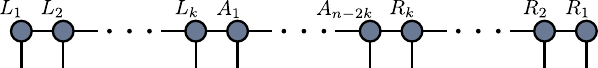} \ ,
\end{equation}
where
\begin{equation}
\begin{split}
\ingrp{1}{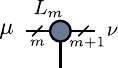} \  & = \
\left[\begin{matrix}
\vert s_m^I \rangle\rangle & \vert s_m^P\rangle\rangle & {} & {} \\
{} & \ddots & \ddots & {} \\
{} & {} & \vert s_m^I\rangle\rangle & \vert s_m^P\rangle\rangle
\end{matrix} \right]_{\mu,\nu} , \\
\ingrp{1}{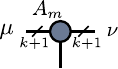} \ & = \ 
\left[\begin{matrix}
\vert s_{k+m}^I\rangle\rangle & \vert s_{k+m}^P\rangle\rangle & {} & {} & {} \\
{} & \ddots & \ddots & {} & {} \\
{} & {} & \vert s_{k+m}^I\rangle\rangle & \vert s_{k+m}^P\rangle\rangle \\
{} & {} & {} & \vert s_{k+m}^I\rangle\rangle
\end{matrix} \right]_{\mu,\nu} , \\
\ingrp{1}{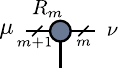} \ &= \ 
\left[\begin{matrix}
\vert s_{n-m+1}^P\rangle\rangle & {} & {}\\
\vert s_{n-m+1}^I\rangle\rangle & \ddots & {} \\
{} & \ddots &  \vert s_{n-m+1}^P\rangle\rangle \\
{} & {} & \vert s_{n-m+1}^I\rangle\rangle
\end{matrix} \right]_{\mu,\nu} ,
\end{split}
\end{equation}
for all $1\leq k \leq \lfloor n/2 \rfloor$. For $\lfloor n/2 \rfloor < k <n$ $\vert \phi_k \rangle\rangle$ can be recovered with the same construction after swapping $\vert s_j^I \rangle\rangle \leftrightarrow \vert s_j^P \rangle\rangle$ and changing $k\gets n - k$.\footnote{Notice that when $n$ is even, the MPS $\vert\phi_{n/2}\rangle\rangle$ has no $A$ tensors.} Lastly, the cases $k=0,n$ are trivially recovered by $\vert \phi_0 \rangle \rangle = \bigotimes_j \vert s_j^I\rangle\rangle$ and $\vert \phi_n \rangle \rangle = \bigotimes_j \vert s_j^P\rangle\rangle$, which can again be written in a simple MPS form with bond dimension $\chi=1$.
\end{lemma}
The proof of this Lemma follows by direct inspection.

\section{MPS for deep HEA circuits composed of U(4) random gates}

It is well known that deep $n$-qubit HEA quantum circuit made of local Haar random i.i.d $U(4)$ gates converge to $2$-designs over the global $U(2^n)$~\cite{brandao2016local,harrow2018approximate}. Using Weingarten calculus we can thus compute the MPS obtained at the end of the $P$-net for these architectures. In particular, if we assume that $U$ forms a  $2$-designs over $U(2^n)$, we have~\cite{mele2023introduction,garcia2023deep}
\small
\begin{equation}\label{eq:2-moment}
\begin{split}
    \tau^{2}_{U}(O^{\otimes 2})=&\frac{1}{d^2 -1}\left(\Tr[O^{\otimes 2}]-\frac{\Tr[O^{\otimes 2}{\rm SWAP}]}{d}\right)\id\otimes \id\\
    &+\frac{1}{d^2 -1}\left(\Tr[O^{\otimes 2}{\rm SWAP}]-\frac{\Tr[O^{\otimes 2}]}{d}\right){\rm SWAP}\,,
\end{split}
\end{equation}
\normalsize
with $d=2^n$. Here 
\begin{equation}\label{eq:defs}
\begin{split}
    \id &= \bigotimes_{j=1}^n I_j\,,\\
    {\rm SWAP} &= \bigotimes_{j=1}^n {\rm SWAP}_{j} = \bigotimes_{j=1}^n \frac{1}{2}(\id_j  + {\rm S}_{j})\,.
\end{split}
\end{equation}
Let us now assume all the Paulis to be normalized, i.e. $\Tr [\id ] = 1$. This leads to $\Tr [\rm S] = 3$, $\Tr [\rm SWAP] = 1/2$, $\Tr [\rm S \, \rm SWAP] = 3/2$.  In what follows, we find it convenient to write 
\begin{equation}\label{eq:vectors}
\begin{split}
    \vert\id \rangle\rangle &= (1, 0) = \ket{0}\,,\\
    \vert {\rm S} \rangle\rangle &= (0, \sqrt{3}) =\sqrt{3}\ket{1}\,\\
    \vert {\rm SWAP} \rangle\rangle &= \frac{1}{2}(\vert\id \rangle\rangle + \vert {\rm S} \rangle\rangle)=\frac{1}{2}(\ket{0}+\sqrt{3}\ket{1}) \,,
\end{split}
\end{equation}
where we have performed a change of basis from $\{\ket{\id}\rangle,\ket{S}\rangle \}$ to $\{\ket{0},\ket{1}\}$ as the latter is orthonormal.

Since we will eventually normalize the vectorized operators to interpret them as quantum states and compute their entropy, we can simply write
\begin{equation}\label{eq:result_of_2-moment}
    \tau^{2}_{U}(O^{\otimes 2}) \propto {\rm SWAP} - \frac{\id}{d} \propto (I+{\rm S})^{\otimes n} - \id \,,
\end{equation}
where we used the fact that Pauli operators are traceless. Vectorizing this operator we get
\begin{equation}\label{eq:vec_result_of_2-moment}
    \vert \tau^{2}_{U}(O^{\otimes 2}) \rangle \rangle \propto (\ket{0} + \sqrt{3}\ket{1})^{\otimes n} - \ket{0}^{\otimes n}\,,
\end{equation}
which once normalized, reads
\begin{equation}\label{eq:ket_result_of_2-moment}
    \ket{\tau^{2}_{U}(O^{\otimes 2})} = \frac{(\ket{0} + \sqrt{3}\ket{1})^{\otimes n} - \ket{0}^{\otimes n}}{\sqrt{d^{2} - 1}}\,.
\end{equation}
Notice that we switched the notation from $\vert \tau^{2}_{U}(O^{\otimes 2}) \rangle \rangle$ to $\vert \tau^{2}_{U}(O^{\otimes 2}) \rangle $ to indicate that the latter is a proper quantum state.

From here, we can compute the reduced density matrix of $\vert \tau^{2}_{U}(O^{\otimes 2}) \rangle $ into any subsystem. In particular, we care for $\rho_{Q_j}=\Tr_{\overline{j}}[\dya{\tau^{2}_{U}(O^{\otimes 2})}]$ where $\Tr_{\overline{j}}$ indicates the partial trace over all qubits but the $j$-th one. In particular, since the state is permutation invariant, $\rho_{Q_j}=\rho_{Q_{j'}}$ for all $j,j'=1,\ldots,n$.  Let us introduce the unnormalized state $\ket{\phi} = \ket{0} + \sqrt{3}\ket{1}$, and let us call $\alpha = 1 / (d^2 - 1)$. We have 
\begin{equation}\label{eq:rho_result_of_2-moment}
\begin{split}
    \dya{\tau^{2}_{U}(O^{\otimes 2})} = \alpha [ &\ketbra{\phi}{\phi}_1\otimes\ketbra{\phi}{\phi}^{\otimes n-1} \\
    & + \ketbra{0}{0}_1\otimes\ketbra{0}{0}^{\otimes n-1}\\
    & - \ketbra{\phi}{0}_1\otimes\ketbra{\phi}{0}^{\otimes n-1}\\ 
    & - \ketbra{0}{\phi}_1\otimes\ketbra{0}{\phi}^{\otimes n-1} ]\,.
\end{split}
\end{equation}
Since $\Tr[\ketbra{\phi}{\phi}^{\otimes n-1}] = 4^{n-1} $, $\Tr[\ketbra{0}{0}^{\otimes n-1}] = \Tr[\ketbra{\phi}{0}^{\otimes n-1}] = \Tr[\ketbra{0}{\phi}^{\otimes n-1}] = 1$, we can straightforwardly read off the reduced density matrix for any single qubit as
\begin{equation}\label{eq:rho_1}
    \rho_{Q_j} = \frac{1}{4^n - 1} \begin{pmatrix}
        4^{n-1} - 1 & \sqrt{3}(4^{n-1} - 1) \\
        \sqrt{3}(4^{n-1} - 1) & 3\,4^{n-1}\\
    \end{pmatrix}\,.
\end{equation}
The eigenvalues $\lambda_1 > \lambda_2$ of this matrix are
\begin{equation}\label{eq:rho1_eigvals}
\begin{split}
    \lambda_1 &= \frac{1}{2} \left(\frac{\sqrt{-5\ 4^n+16^n+13}}{4^n-1}+1\right)\,,\\
    \lambda_2 &= \frac{1}{2}-\frac{\sqrt{-5\ 4^n+16^n+13}}{2 \left(4^n-1\right)}\,.
\end{split} 
\end{equation}
Interestingly, $\lambda_2 \neq 0$, signaling that the ensuing MPS for deep circuits has a non-trivial entanglement structure (it is not a product state). However, these correlations become exponentially small as the system size $n$ increases.

\section{Derivation of the $O(4)$ $P$-gates}\label{app:so4_pgate}

In the same spirit as the toy model presented in the main text for $U(4)$, we here show how to build the $P$-matrices of random circuits whose gates are sampled i.i.d. from the Haar measure of the fundamental representation of $O(4)$. We start by recalling that ${\rm comm}^{(2)}(O(4))$ is~\cite{garcia2023deep}:
\begin{equation}
    {\rm comm}^{(2)}(O(4))={\rm span}_{\mathbb{C}}\{\id\otimes\id, {\rm SWAP}, \Pi \}\nonumber\,,
\end{equation}
where $\Pi$ is the unnormalized projector $\Pi=d\ketbra{\Phi}{\Phi}$ onto $\ket{\Phi} = \frac{1}{2}\sum_{i=0}^{3}\ket{ii}$, the Bell state over two copies of $\HC$, the (two-qubit) Hilbert space $\HC = \HC_1 \otimes \HC_2$ over which a gate sampled from $O(4)$ acts on. We now notice that, analogously to the case of the ${\rm SWAP}$ operator, $\Pi$ can be decomposed as $\Pi=\Pi_1\otimes \Pi_2$, with $\Pi_j$ acting on $\HC_j^{\otimes 2}$, the two copies of the $j$-th qubit, and that 
\begin{equation}
    \Pi = \frac{1}{2}(\id + {\rm B}) \nonumber \,,
\end{equation}
with ${\rm B}=X\otimes X- Y\otimes Y+Z\otimes Z$  and $X,Y$ and $Z$ the Pauli operators. With this in mind, we find it convenient to work with the following basis of ${\rm comm}^{(2)}(O(4))$:
\begin{equation}
    \{\id\otimes\id, {\rm S}\otimes\id+\id\otimes {\rm S}+{\rm S}\otimes {\rm S}, {\rm B}\otimes\id+\id\otimes {\rm B}+{\rm B}\otimes {\rm B} \}\nonumber\,,
\end{equation}
where we recall the definition ${\rm S}=X\otimes X+Y\otimes Y+Z\otimes Z$ from the main text.
Let us now define the following two vectors
\begin{align}
 |A_{\rm S}\rangle\rangle &= (|{\rm S}\rangle\rangle|\id\rangle\rangle+|\id\rangle\rangle|{\rm S}\rangle\rangle+|{\rm S}\rangle\rangle|{\rm S}\rangle\rangle)\,,\nonumber\\
 |A_{\rm B}\rangle\rangle &= (|{\rm B}\rangle\rangle|\id\rangle\rangle+|\id\rangle\rangle|{\rm B}\rangle\rangle+|{\rm B}\rangle\rangle|{\rm B}\rangle\rangle)\,.\nonumber
\end{align}
Given these, we can express the vectorized basis of ${\rm comm}^{(2)}(O(4))$ as
\begin{equation}
    \left\{|\id\rangle\rangle|\id\rangle\rangle,|A_{\rm S}\rangle\rangle, |A_{\rm B}\rangle\rangle\right\}\,.
\end{equation}
We can now compute the action of the second moment operator $\widehat{\tau}_{O(4)}^{(2)}$ on each distinct tensor product of two elements from $\{|\id\rangle\rangle, |{\rm S}\rangle\rangle, |{\rm B}\rangle\rangle\}$. We find 
\begin{align}
\widehat{\tau}_{SO(4)}^{(2)}|\id\rangle\rangle|\id\rangle\rangle&=|\id\rangle\rangle|\id\rangle\rangle\nonumber\,,\\
\widehat{\tau}_{SO(4)}^{(2)}|{\rm S}\rangle\rangle|\id\rangle\rangle &= \widehat{\tau}_{SO(4)}^{(2)}|{\id}\rangle\rangle|{\rm S}\rangle\rangle =\frac{7}{36}A_2 + \frac{1}{36} A_3 \nonumber\,,\\
\widehat{\tau}_{SO(4)}^{(2)}|{\rm S}\rangle\rangle|{\rm B}\rangle\rangle &= \widehat{\tau}_{SO(4)}^{(2)}|{\rm B}\rangle\rangle|{\rm S}\rangle\rangle =\frac{1}{6}A_2 + \frac{1}{6} A_3 \nonumber\,,\\
\widehat{\tau}_{SO(4)}^{(2)}|{\rm S}\rangle\rangle|{\rm S}\rangle\rangle&=\frac{11}{18}A_2 - \frac{1}{18} A_3 \nonumber\,,\\
\widehat{\tau}_{SO(4)}^{(2)}|{\rm B}\rangle\rangle|{\id}\rangle\rangle &= \widehat{\tau}_{SO(4)}^{(2)}|{\id}\rangle\rangle|{\rm B}\rangle\rangle=\frac{1}{36}A_2 + \frac{7}{36} A_3 \nonumber\,,\\
\widehat{\tau}_{SO(4)}^{(2)}|{\rm B}\rangle\rangle|{\rm B}\rangle\rangle&=\frac{-1}{18}A_2 + \frac{11}{18} A_3 \nonumber\,.
\end{align}

Putting it all together, the previous shows that the $P$-matrix for the fundamental representation of $O(4)$ is
\begin{equation}\label{eq:matrix-P-so-0-ap}
     P=\begin{pmatrix}  
1 & 0 & 0 & 0 & 0 & 0 & 0 & 0 & 0 \\
0 & {7/36} & {1/36} & {7/36} & {11/18} & {1/6} & {1/36} & {1/6} & -{1/18} \\
0 & {1/36} & {7/36} & {1/36} & -{1/18} & {1/6} & {7/36} & {1/6} & {11/18} \\
0 & {7/36} & {1/36} & {7/36} & {11/18} & {1/6} & {1/36} & {1/6} & -{1/18} \\
0 & {7/36} & {1/36} & {7/36} & {11/18} & {1/6} & {1/36} & {1/6} & -{1/18} \\
0 & 0 & 0 & 0 & 0 & 0 & 0 & 0 & 0 \\
0 & {1/36} & {7/36} & {1/36} & -{1/18} & {1/6} & {7/36} & {1/6} & {11/18} \\
0 & 0 & 0 & 0 & 0 & 0 & 0 & 0 & 0 \\
0 & {1/36} & {7/36} & {1/36} & -{1/18} & {1/6} & {7/6} & {1/6} & {11/18}
\end{pmatrix}\,.\nonumber
\end{equation}
with $P$ acting on two, three-dimensional vector spaces with basis $\{|\id\rangle\rangle, |{\rm S}\rangle\rangle, |{\rm B}\rangle\rangle\}$.

\clearpage

\end{document}